\documentclass[journal]{IEEEtran}
\makeatletter
\usepackage{graphicx,cite,calc,xcolor,subfigmat,amssymb,amsmath,mathrsfs,epstopdf}

\usepackage{verbatim}
\usepackage{graphicx}
\usepackage{caption}
\usepackage{epstopdf}
\usepackage{amsmath,amssymb,amsfonts,bm}
\usepackage{subfigure}
\usepackage{psfrag}
\usepackage[dvips]{epsfig}
\usepackage{amsthm}
\usepackage{color}
\usepackage[usenames,dvipsnames]{pstricks}
\usepackage{epsfig}
\usepackage{subfigure}
\usepackage{color}
\usepackage{cite}
\usepackage{tikz}
\usetikzlibrary{shapes,arrows,automata}
\newtheorem{thm}{Theorem}
\newtheorem{lemm}{Lemma}
\newtheorem{prp}{Proposition}
\newtheorem{cor}{Corollary}
\newtheorem{rema}{Remark}

\DeclareMathOperator{\tr}{tr}

\DeclareMathOperator{\vect}{vec}

\def\bx{{\bf x}}

\def\bn{{\bf n}}

\def\bZ{{\bf Z}}
\def\bD{{\bf D}}
\def\by{{\bf y}}
\def\bY{{\bf Y}}
\def\bV{{\bf V}}
\def\bW{{\bf W}}
\def\bX{{\bf X}}

\def\bb{{\bf b}}
\def\bH{{\bf H}}

\def\bQ{{\bf Q}}

\def\bC{{\bf C}}
\def\bB{{\bf B}}

\def\bG{{\bf G}}

\def\bI{{\bf I}}

\def\bd{{\bf d}}

\def\bA{{\bf A}}

\def\bU{{\bf U}}

\def\bs{{\bf s}}

\def\bS{{\bf S}}

\def\complexC{{\mathbb{C}}}
\def\realR{{\mathbb{R}}}
\def\realS{{\mathbb{S}}}
\def\bx{{\bf x}}

\def\bF{{\bf F}}
\def\bQ{{\bf Q}}

\def\bTheta{
{\boldsymbol \Theta}}

\def\bPhi{{\bf \Phi}}

\def\bOmega{{\bf \Omega}}

\def\bOmega{{\bf \Omega}}
\def\bGamma{{\bf \Gamma}}

\def\bzero{{\boldsymbol 0}}

\def\Fig{Fig. }
\begin{document}

This paper is accepted for publication in IEEE Transactions on Signal Processing with DOI 10.1109/TSP.2019.2929470. IEEE copyright notice.
© 2019 IEEE. Personal use of this material is permitted. Permission from IEEE must be obtained
for all other uses, in any current or future media, including reprinting/republishing this material
for advertising or promotional purposes, creating new collective works, for resale or redistribution
to servers or lists, or reuse of any copyrighted.
\newpage

\title{Max-Min Fairness Design for MIMO Interference Channels: a Minorization-Maximization Approach}

\author{Mohammad Mahdi Naghsh$^*$, \emph{Member, IEEE}, Maryam Masjedi, Arman Adibi, and Petre Stoica,  \emph{Fellow, IEEE}

\thanks{M. M. Naghsh, M. Masjedi and A. Adibi are with the Department of Electrical and Computer Engineering, Isfahan University of Technology, Isfahan 84156-83111, Iran. P. Stoica is with the Department of Information Technology, Uppsala University, Uppsala, SE 75105, Sweden.

This work was supported by Iran National Science Foundation (INSF) and Isfahan  University of Technology (Gant No. 96006628).

 *Please address all the correspondence to M. M. Naghsh, Phone: (+98) 31-33912450; Fax: (+98) 31-33912451; Email: mm\_naghsh@cc.iut.ac.ir} }

%\thanks{* Please address all the correspondence to M. M. Naghsh, Phone: (+98) 31-33912450; Fax: (+98) 31-33912451; Email: mm\_naghsh@cc.iut.ac.ir}

\maketitle

\begin{abstract}
We address the problem of linear precoder (beamformer) design in a multiple-input multiple-output interference channel (MIMO-IC). The aim is to design the transmit covariance matrices in order to achieve  max-min utility fairness for all users.  The corresponding optimization problem is non-convex and NP-hard in general. We devise  an efficient  algorithm based on the minorization-maximization (MM) technique to obtain quality solutions to this problem. The proposed method solves a second-order cone convex program (SOCP) at each iteration. We prove that the devised method converges to  stationary points of the problem. We also extend our algorithm to the case where there are uncertainties in the noise covariance matrices or channel state information (CSI).  Simulation results show the effectiveness of the proposed method compared with its main competitor.
\end{abstract}

Keywords: Interference channel, Minorization-maximization (MM), Max-min fairness, MIMO,  Rate optimization.

\section{Introduction}
We consider the linear precoder design problem in a MIMO interference channel in which a set of transmitter-receiver pairs communicate over a shared (time or frequency) resource. The precoder matrices can be designed to improve the network performance from a sum rate or  minimum rate (max-min fairness) point of view \cite{zander1992performance,155977,1634798,5638157,1262126,679579,1561584,liu2011max,
razaviyayn2011linear,razaviyayn2013linear,aquilina2017weighted,huberman2015mimo,wang2017upper,zhang2017sum,razaviyayn2012linear,naghsh2016efficient,liu2017dynamic}.

The problem of linear transceiver design under the max-min fairness criterion has been widely studied in the literature \cite{zander1992performance,155977,1634798,5638157,1262126,679579,1561584,liu2011max,razaviyayn2011linear,razaviyayn2013linear}. In \cite{zander1992performance} and \cite{155977}, the power control problem under a max-min signal-to-interference-plus-noise ratio (SINR) criterion has been studied and performance bounds for power control algorithms have been obtained. The problem of designing the transmitter precoder that maximizes the minimum rate of users in a multiple-input single-output (MISO) network is also studied in \cite{1634798,679579,5638157,1262126}.
The authors of \cite{1561584} maximized the worst case SINR  subject to a power constraint on the design precoder matrices in a MIMO-IC and showed this problem can be solved using standard conic optimization packages. The authors of \cite{liu2013max} considered the max-min fairness precoder design in a single-input multiple-output (SIMO) IC and showed that this problem can be solved in polynomial time. In \cite{liu2011max}, the authors recast the max-min fairness problem in MIMO-IC as the problem of  finding the globally optimal transceiver that maximizes the minimum SINR among all users. They showed that when each transmitter (receiver) is equipped with more than one antenna and each receiver (transmitter) is equipped with more than two antennas, the problem is strongly NP-hard. To deal with the problem they  proposed two algorithms which decompose the original NP-hard problem into a series of convex subproblems. The authors of \cite{liu2017dynamic} further showed that the max-min fairness problem in MIMO-IC is strongly NP-hard when each transmitter and receiver is equipped with more than one antenna.
In \cite{razaviyayn2011linear} and \cite{razaviyayn2013linear}, the authors considered the problem of linear precoder design for MIMO-IC under a max-min fairness criterion and showed that  when there are at least two antennas at each transmitter and receiver, the problem belongs to a class of NP-hard problems. They proposed an  algorithm that computes an approximate solution to the original problem. Note that in the aforementioned works, the precoder matrices are designed for the cases in which the number of symbols in a stream is assumed to be a priori known.

\begin{table}[t]
	\caption{Notations} \label{table:notations}
	\centering
	\begin{tabular}{ll}
		\hline \hline
		$\| \bx \|_n$: & the $l_n$-norm of the vector $\bx$, defined as $\left( \sum_k |x(k)|^n \right)^\frac{1}{n}$ \\
        $\| \bX \|_2$: & the spectral norm of the matrix $\bX$ i.e. the largest singular value of $\bX$\\
		$ \bX^H$: & the  conjugate transpose of  matrix $\bX$ \\
		tr($\bX$): & the trace of  matrix $\bX$ \\
$\lambda_{max}(\bX)$: & the maximum eigenvalue of hermitian matrix $\bX$\\
				$\bA \otimes \bB$: & the Kronecker product of two matrices $\bA$ and $\bB$ \\
$\bX \succeq \bY$: & $\bX-\bY$ is positive semidefinite  \\
$\bX \succ \bY$: & $\bX-\bY$ is positive definite  \\
$\bX^{\frac{1}{2}}$: & the Hermitian square root of the positive semidefinite matrix $\bX$\\
                     & i.e. $\bX=\bX^{\frac{1}{2}}(\bX^{\frac{1}{2}})^H$\\
$\mbox{vec}(\bX)$: &the vector obtained by column-wise stacking of $\bX$\\
$\bI_n$: &the identity matrix of $\complexC^{n \times n}$\\
		$\realR$: & the set of real numbers \\
		$ \complexC$: & the set of complex numbers \\
$ \Re(x)$: & the real part of $x$ \\
$\realR_{+}$: & the set of nonnegative real numbers \\
$\realS^{+}_N$: & the set of positive semidefinite matrices of $\complexC^{N \times N}$\\
$\realS^{++}_N$: & the set of positive definite matrices of $\complexC^{N \times N}$\\

		\hline \hline
		\end{tabular}
		\end{table}

The precoder design  for achieving max-min rate fairness among users in MIMO-IC leads to a non-convex and, in general, NP-hard problem. Some works \cite{razaviyayn2011linear}\cite{razaviyayn2013linear} address and tackle this design problem by using block coordinate descent as an optimization technique. At the same time, minorization-maximization (MM)\footnote{Also known as MaMi or MiMa in the literature \cite{naghsh2016efficient}.}, a general iterative  optimization technique which is often quiet stable and shown to be difficult to be outperformed \cite{hunter2004mm}\cite{hunter2004tutorial}\cite{sun2017majorization}, has recently been successfully employed to deal with several non-convex problems in communication/active sensing systems, see e.g. \cite{naghsh2016efficient}\cite{naghsh}. In light the good properties of MM technique,  we consider using it for precoder design in MIMO-IC. The main contributions of the present paper can be summarized as follows:
\begin{itemize}
  \item We design the transmit covariance matrices (the number of transmitted symbols is not necessarily given) under a max-min fairness criterion for systems using the conventional linear  minimum mean square error (LMMSE) receivers. We propose an efficient algorithm based on the MM  technique to obtain quality solutions to this design problem.

      \item We prove that the obtained solutions are stationary points of the problem. This result is obtained as a corollary of a general theorem that can be used to analyze the convergence of  MM algorithms for an entire class of maxmin optimization problems.%, we propose a theorem that can also be used for convergence analysis of a class of maxmin problems tacking with MM algorithm.

  \item Compared with \cite{razaviyayn2011linear} and \cite{razaviyayn2013linear}, we consider a more general case of designing the precoder covariance matrices, which means that the optimal number of symbols in a stream is also obtained as a by-product.

  \item We also extend our algorithm to the practical cases where there are uncertainties in the noise
covariance matrices or in the CSI.
\end{itemize}

The rest of the paper is organized as follows. The signal and system model along with the associated max-min precoder covariance design problem are described in Section II. The proposed method for designing the precoder covariances as well as the  precoder matrices  under the max-min fairness criterion is derived in Section III. This section also includes a convergence analysis of the proposed method. Precoder design under noise covariance uncertainty and imperfect CSI is considered in Section IV. Numerical results are provided in Section V and, finally, conclusions are drawn in Section VI.

Table I summarizes the notation used throughout this paper.

\begin{figure}[t]
\centerline{{\resizebox{!}{.6\columnwidth}{\includegraphics{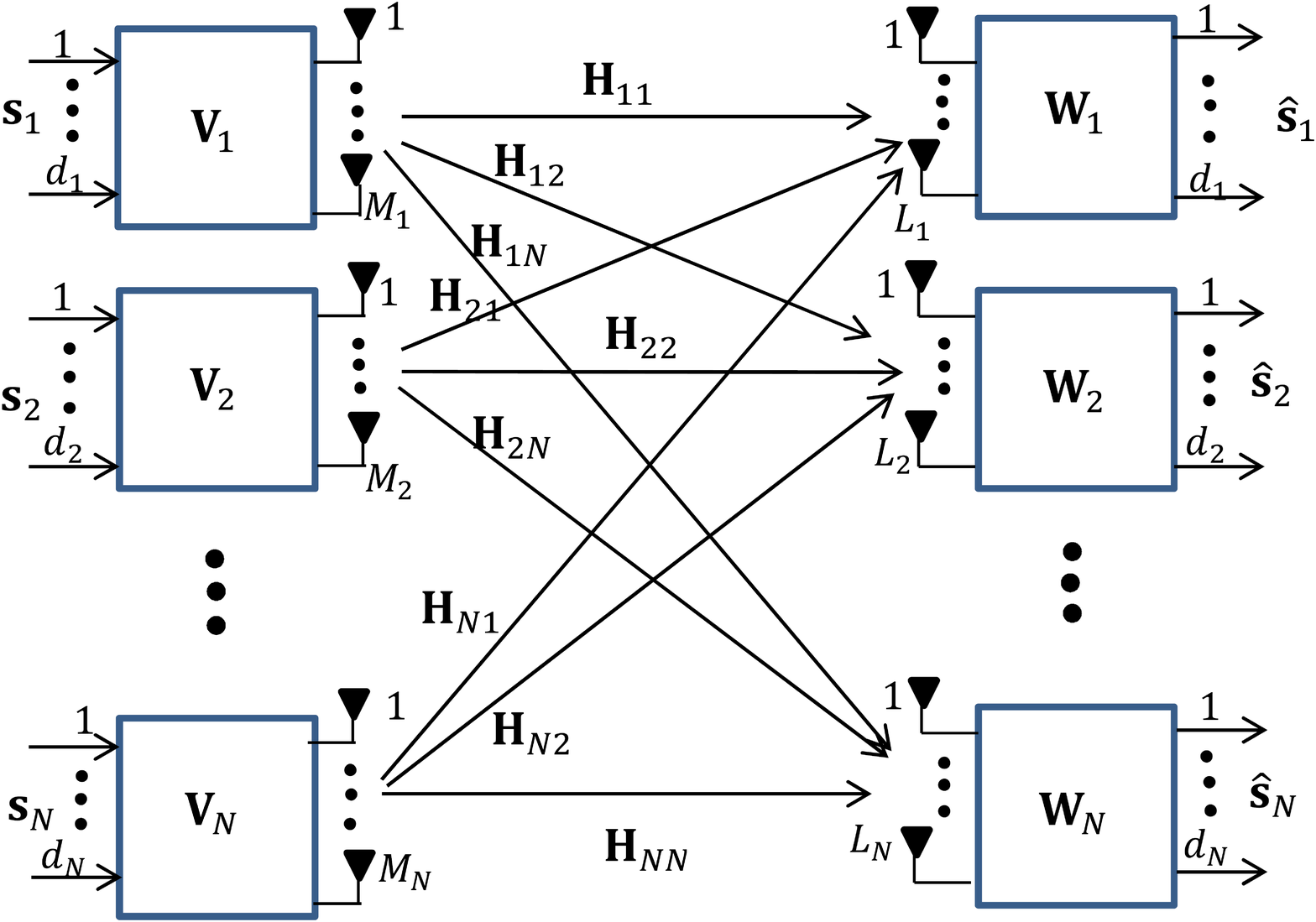}}}}

\caption{A generic MIMO-IC.}\label{system}
\end{figure}

\section{System model and Problem formulation}
Consider $N$ transmit-receive pairs communicating over a MIMO interference channel as shown in Fig.~\ref{system}.
We assume that the $i$th transmitter and the $j$th receiver are equipped with $M_i$ and $L_j$ antennas, respectively. The $i$th transmitter uses the linear precoder matrix $\bV_i\in \complexC^{M_i\times d_i}$ to convert the symbol stream $\bs_i\in \complexC^{d_i\times 1}$ (consisting of $d_i$ independent data symbols) into the vector $\bd_i\in \complexC^{M_i\times 1}$, i.e.,
\begin{equation}\label{t_beamformer}
  \bd_i=\bV_i \bs_i
\end{equation}
and sends it over flat fading channels. The received signal at the $i$th receiver is given by:
\begin{equation}\label{model}
  \by_i=\underbrace{\bH_{ii}\bd_i}_{\textrm{desired signal}}+\underbrace{\sum_{j\neq i}\bH_{ji}\bd_j+\bn_i}_{\textrm{interference plus noise}}
\end{equation}
where $\bH_{ji}\in \complexC^{L_i\times M_j}$ denotes the channel matrix between the $j$th transmitter and the $i$th receiver. Also, $\bn_i\in \complexC^{L_i\times 1}$ is the circularly symmetric complex Gaussian (CSCG) noise at the $i$th receiver with zero mean and covariance matrix $\bGamma_i \in \realS^{++}_{L_i}$.  The $i$th receiver uses the linear decoder matrix $\bW_i\in \complexC^{d_i\times L_i}$ to obtain $\hat{\bs}_i\in \complexC^{d_i\times 1}$ which is an estimate of the transmitted vector $\bs_i$:
\begin{align}\label{r_beamformer}
  \hat{\bs}_i= & \bW_i \by_i \\ \nonumber
  = & \bW_i \bH_{ii} \bV_i \bs_i + \bW_i \sum_{j\neq i}\bH_{ji} \bV_j \bs_j + \bW_i \bn_i
\end{align}

 Assuming the symbol stream $\bs_i$ is a Gaussian random vector with zero mean and covariance matrix $\bI_{d_i}$, the rate of the $i$th user is given by \cite{negro2010mimo}:
\begin{equation}\label{Rate1}
  R_i  =  \log \det \left(\bI_{d_i} + \bW_i \bH_{ii} \bV_i \bV_i^H \bH_{ii}^H \bW_i^H \left(\bW_i \bC_{\overline{i}} \bW_i ^H\right)^{-1}\right)
\end{equation}
with $\bC_{\overline{i}}$ being the interference plus noise covariance matrix  defined as
\begin{equation}\label{IPN}
  \bC_{\overline{i}}  = \bGamma_i + \sum_{j\neq i} \bH_{ji} \bV_j \bV_j^H \bH_{ji}^H
\end{equation}

Employing the conventional  LMMSE decoder at the receivers means that the $i$th decoder matrix is given by
%It is well known that, among all linear precoders, the LMMSE decoder is optimal for interference suppression \cite{negro2010mimo}. Therefore, we assume that the receivers employ the LMMSE decoder, viz.,
\begin{equation}\label{LMMSE}
  \bW_i^{\textrm{LMMSE}}=\bV_i^H \bH_{ii}^H\left(\sum_{j=1}^N \bH_{ji} \bV_j \bV_j^H \bH_{ji}^H + \bGamma_i \right)^{-1}
\end{equation}

By substituting \eqref{LMMSE} into \eqref{Rate1}, it can be  verified that (for completeness we include a proof of (\ref{LMMSE}) and (\ref{Rate}) in Appendix A):
\begin{equation}\label{Rate}
  R_i  =  \log \det \!\left(\!\!\bI_{d_i}+\bV_{i}^{H}\bH_{ii}^{H}\!\left[\bGamma_i+  \sum_{j \neq i}^{} \bH_{ji}\bV_{j}\bV_{j}^H\bH_{ji}^H\right]^{-1}\!\!\!\!\bH_{ii}\bV_{i}\!\!\right)
\end{equation}

 \rema{\normalfont { The system model and the proposed design methodology in this paper can be extended to the MIMO interference broadcast channel (MIMO-IBC) (see Appendix B for details on the MIMO-IBC case). }

\rema{\normalfont {Interestingly, using the decoder  ${\bW}_i^{\prime}=\bV_i^H \bH_{ii}^H\bC_{\overline{i}}^{-1}$, see (\ref{IPN}), leads to the same rate as the LMMSE, see (\ref{Rate}). Furthermore, the matrix ${\bW}_i^{\prime}$ maximizes the rate in (\ref{Rate1}). To see this, use standard properties of Schur complement to verify that the inequality
\begin{equation}\label{schuar1}
  \bV_i^H\bH_{ii}^H\bW_i^H (\bW_i \bC_{\overline{i}} \bW^H)^{-1} \bW_i \bH_{ii}\bV_i \preceq \bV_i^H \bH_{ii}^H \bC_{\overline{i}}^{-1} \bH_{ii}\bV_i
\end{equation}
is equivalent to the positive semi-definitness of the matrix:
\begin{equation}\label{schur2}
\bPhi_i=\left[
        \begin{array}{cc}
          \bV_i^H \bH_{ii}^H \bC_{\overline{i}}^{-1} \bH_{ii}\bV_i & \bV_i^H\bH_{ii}^H\bW_i^H \\
          \bW_i \bH_{ii}\bV_i & \bW_i \bC_{\overline{i}} \bW^H \\
        \end{array}
        \right].
\end{equation}
Now, observe that the matrix $\bPhi_i$ above indeed is in $\realS^{+}_{d_i}$ because it can be decomposed as $\bPhi_i=\bTheta_i \bTheta_i^H$ with
 \begin{equation}\label{schur3}
\bTheta_i=\left[
        \begin{array}{cc}
          \bV_i^H \bH_{ii}^H  & \bzero \\
          \bzero & \bW_i \\
        \end{array}
        \right]
        \left[
        \begin{array}{cc}
          \bC_{\overline{i}}^{-1/2}  \\
          \bC_{\overline{i}}^{1/2}  \\
        \end{array}
        \right]
       \end{equation}
Therefore, (\ref{schuar1}) holds true. Moreover, it can be  verified that by substituting $\bW_i=\bW_i^{\prime}=\bV_i^H \bH_{ii}^H\bC_{\overline{i}}^{-1}$ in (\ref{schuar1}), the left-hand side  becomes $\bV_i^H \bH_{ii}^H \bC_{\overline{i}}^{-1} \bH_{ii}\bV_i$ which is equal to the right-hand side. Therefore, $\bW_i^{\prime}$ maximizes the rate in (\ref{Rate1}). Because the LMMSE decoder in (\ref{LMMSE}) and $\bW_i^{\prime}$  yield the same rate, we  conclude that the LMMSE decoder maximizes the rate as well. Note that the optimality of this decoder for mean square error minimization has been addressed in \cite{5756489,razaviyayn2013linear} (see also \cite{schmidt2009minimum}).\hfill $\blacksquare$
}

 Using  Sylvester's determinant property, i.e. $\det(\bI+\bA \bB)=\det(\bI+\bB \bA)$, the rate $R_i$ in \eqref{Rate} can be rewritten as
 \begin{equation}\label{Rate2}
  R_i  =  \log \det \!\left(\bI_{L_i}+\bH_{ii} \bQ_i \bH_{ii}^{H}\!\left[\bGamma_i+ \sum_{j \neq i}^{} \bH_{ji} \bQ_j \bH_{ji}^H\right]^{-1}\right)
\end{equation}
 where $\bQ_i \triangleq \bV_{i}\bV_{i}^{H} \in \complexC^{M_i\times M_i}, i=1,\dots , N$, are the precoder covariance matrices. In this paper, the goal is to design the precoder covariance matrices $\{\bQ_i\}_{i=1}^N$ to maximize the minimum rate of the users, which can be cast as the following problem:
\begin{align}\label{p}
\!\max_{\{\bQ_{i}\}_{{i=1}}^N}\, \min_{i=1,2,\dots , N} \quad & {R_i} \\ \nonumber
 \text{s.t.} \qquad & \tr\{\bQ_i\} \leq  p_{i} \qquad & \forall i=1,2,\dots , N \\ \nonumber
& \bQ_i \succeq \textbf{0} \qquad &\forall i=1,2,\dots , N \nonumber
\end{align}
where $p_{i}$ is the power available to the $i$th transmitter.
Note that in the covariance design approach, we jointly design the optimum precoder matrices $\{\bV_i\}_{i=1}^{N}$, as well as, the optimum number of their columns $\{d_i\}_{i=1}^{N}$, i.e. the length of symbol streams. More precisely, we fully exploit the available degrees of freedom of the design problem instead of considering the design problem in a limited framework in which $\{d_i\}_{i=1}^{N}$ are assumed to be a priori known.  %\footnote{Note that herein we exploit the available degrees of freedom and hence, the obtained solutions can be implemented. This should not be confused with e.g. semi-definite relaxation (SDR) in which the problem is solved in a space with higher dimension than the original problem. The solution obtained by SDR usually cannot be implemented in practice, viz. is infeasible, and obtaining a  feasible solution is associated with a synthesis loss. }.

In the next section, we assume that the noise covariance matrices ${\{\bGamma_i\}}_{i=1}^N$ as well as the channel matrices $\{\bH_{ij}\}_{{i,j=1}}^N$ are exactly known. We consider the case of uncertain a priori knowledge in Section IV.

\section{the proposed method}
\subsection{Derivation of the proposed method}
It can be shown that the design problem in \eqref{p} is non-convex and NP-hard in general \cite{razaviyayn2013linear}. In what follows we devise a method based on the minorization-maximization (MM) technique \cite{stoica2004cyclic} to tackle this problem.

In \eqref{p}  the constraints are convex but the objective function is non-convex.
Therefore we will apply the MM technique to the objective function. For this purpose, we first introduce the following proposition.
\prp{
\normalfont The rate $R_i$, see \eqref{Rate2}, can be rewritten as:
\begin{equation}\label{R1}
  R_i=\log \det (\bU^{H}\bB_{i}^{-1}\bU)
\end{equation}
where $\bU$ and $\bB_{i}$ are defined as,
\begin{equation}\label{U}
  \bU\triangleq\left[
                                                          \begin{array}{cc}
                                                            \bI_{M_i} & \textbf{0}_{M_i\times L_i} \\
                                                          \end{array}
                                                        \right]^T
\end{equation}
and
\begin{equation}\label{bi}
\bB_{i}=\left[
        \begin{array}{cc}
          \bI_{M_i} & \widetilde{\bV}^{H}_i\bH_{ii}^{H} \\
          \bH_{ii}\widetilde{\bV}_i & \bGamma_i+\displaystyle\sum_{j=1}^{N}\bH_{ji} \widetilde{\bV}_j\widetilde{\bV}_j^{H}\bH_{ji}^{H} \\
        \end{array}
        \right]
\end{equation}
with $\widetilde{\bV}_i\triangleq \bQ_{i}^{\frac{1}{2}}\in \complexC^{M_i\times M_i}$.

 \IEEEproof See Appendix C. \hfill $\blacksquare$}
\normalfont
\\
By using  \eqref{R1}, the problem in \eqref{p} can be rewritten as follows
\begin{align}\label{pmin}
   \max_{\{\widetilde{\bV}_i\}_{{i=1}}^N,}& \min_{i=1,\cdots, N}  \log \det (\bU^{H}\bB_{i}^{-1}\bU)\ \\ \nonumber
\text{s.t.} \quad &
 \tr\{\widetilde{\bV}_i\widetilde{\bV}_i^H\} \, \leq \,  p_{i}, \,\,\,  \forall i=1,2,\dots , N\\ \nonumber
\end{align}
The following lemma  (see, e.g., \cite{naghsh})  lays the ground for applying MM to \eqref{pmin}.

\lemm{
\normalfont The function $f(\bX)=\log\det(\bU^H\bX^{-1}\bU)$: $\realS^{++}_N\rightarrow\realR_{+}$ is convex for any full column rank matrix $\bU. $
\hfill $\blacksquare$}
\normalfont
\\

Using Lemma 1 and noting that $\bB_i\succ \bf{0}$, $\forall i=1,2,\dots,N$ (see Appendix D), the objective function in  problem \eqref{pmin} can be minorized at a given $\overline{\bB}_{i}$ as follows
\begin{align} \label{hyp}
  \log \det (\bU^{H}\bB_{i}^{-1}\bU)\,\geq\,&\log \det  (\bU^{H}\overline{\bB}_{i}^{-1}\bU) \\ \nonumber
  &- \tr\{\bF_{i}(\bB_{i}-\overline{\bB}_{i})\}
\end{align}
where $\bF_i$ is given by (see Appendix E):
\begin{equation}\label{F}
   \bF_{i}=\overline{\bB}_{i}^{-1}\bU(\bU^{H}\overline{\bB}_{i}^{-1}\bU)^{-1}\bU^{H}\overline{\bB}_{i}^{-1}.
\end{equation}
Note that $\overline{\bB}_{i}$ can be chosen as the value of ${\bB}_{i}$ at the $(\kappa-1)$th iteration. Consequently, let
\begin{align}\label{gi}
  g_i^{(\kappa)}(\widetilde{\bV}_{1},\cdots,\widetilde{\bV}_{N})&\triangleq \log \det  (\bU^{H}{({\bB}^{(\kappa-1)}_{i})}^{-1}\bU) \\ \nonumber
  &- \tr\{\bF_{i}(\bB_{i}-{\bB}^{(\kappa-1)}_{i})\}
\end{align}
 (we omit the dependence of $\bF_i$ on the iteration number to simplify the notation). Then it follows from \eqref{hyp} that the objective function in  \eqref{pmin} can be minorized at the $\kappa$th iteration by:
\begin{align} \label{mnz}
 \min_{i=1,\cdots, N} \log \det (\bU^{H}\bB_{i}^{-1}\bU)\,\geq\, \min_{i=1,\cdots, N} g_i^{(\kappa)}(\widetilde{\bV}_{1},\cdots,\widetilde{\bV}_{N})
\end{align}
The MM technique that makes use of \eqref{mnz}, consists of  iteratively solving the following problem (for $\kappa=1,2,...$):
\begin{align}\label{p2prime1}
\max_{\{\widetilde{\bV}_{i}\}_{{i=1}}^N} & \min_{i=1,\cdots, N} \quad g_i^{(\kappa)}(\widetilde{\bV}_{1},\cdots,\widetilde{\bV}_{N}) \\ \nonumber
\text{s.t.} \quad & \tr\{\widetilde{\bV}_{i}\widetilde{\bV}_{i}^H\} \, \leq \,  p_{i}, \,\,\, \forall i=1,2,\dots , N
\end{align}

Next, we rewrite (\ref{p2prime1})  using an auxiliary variable $t$:
\begin{align}\label{p2prime}
\max_{\{\widetilde{\bV}_{i}\}_{{i=1}}^N,t}&  \quad t \\ \nonumber
\text{s.t.} \quad & g_i^{(\kappa)}(\widetilde{\bV}_{1},\cdots,\widetilde{\bV}_{N})\geq t,  \,\,\,& \forall i=1,2,\dots , N \\ \nonumber
&\tr\{\widetilde{\bV}_{i}\widetilde{\bV}_{i}^H\} \, \leq \,  p_{i}, \,\,\, & \forall i=1,2,\dots , N
\end{align}

To derive an explicit expression for the constraints $g_i^{(\kappa)}(\widetilde{\bV}_{1},\cdots,\widetilde{\bV}_{N})\geq t$ in terms of the design variables $\{\widetilde{\bV}_{i}\}_{{i=1}}^N$, let
\begin{equation}\label{F_def}
  \bF=\left(
                          \begin{array}{cc}
                            \bF_{{11}_{M_i\times M_i}} & \bF_{{12}_{M_i\times L_i}} \\
                            \bF_{{21}_{L_i\times M_i}} & \bF_{{22}_{L_i\times L_i}} \\
                          \end{array}
                        \right).
\end{equation}
Then,  combining \eqref{F_def} and  \eqref{bi}, it can be verified  that:
\begin{align}\label{Lemma2}
   \tr{\{\bF_{i}\bB_{i}\}} =&  2 \Re\{\tr\{(\bF_i)_{12} \bH_{ii}\widetilde{\bV}_{i} \}\} +\tr\{(\bF_i)_{11}\} \\ \nonumber
    & +\tr\{(\bF_i)_{22}\bGamma_i\}\\ \nonumber
    &+\tr\{(\bF_i)_{22}\sum_{j=1}^{N}\bH_{ji}\widetilde{\bV}_{i}\widetilde{\bV}_{i}^H\bH_{ji}^H\}.\nonumber
  \end{align}

\normalfont Defining $\bx_i\triangleq\vect(\widetilde{\bV}_{i})$ and using properties of the vectorization operator, viz. $\tr(\bA\bB)=(\vect(\bA^T))^T\vect(\bB)$ and $\vect(\bA\bB\bC)=(\bC^T\otimes \bA)\vect(\bB)$ (for any arbitrary matrices $\bA$, $\bB$ and $\bC$), we can rewrite the first and the last terms in \eqref{Lemma2} as:
\begin{align}
  &\tr\{(\bF_i)_{12}\bH_{ii}\widetilde{\bV}_{i} \}  = \bb_i^H\bx_i \label{gb}\\
  &\tr\{(\bF_i)_{22}\sum_{j=1}^{N}\bH_{ji}\widetilde{\bV}_{i}\widetilde{\bV}_{i}^H\bH_{ji}^H\}  =\sum_{j=1}^{N}\bx_j^H\bG_{ji}\bx_j \label{gb1}
\end{align}
where $\bb_i\triangleq\vect(\bH_{ii}^H(\bF_i)_{12}^H)$ and $\bG_{ji}\triangleq\bI_M\otimes(\bH_{ji}^H(\bF_i)_{22}\bH_{ji})$. Note that according to the Kronecker product properties, $\bG_{ji}\succeq \textbf{0} $, because $\bH_{ji}^H(\bF_i)_{22}\bH_{ji}$ is positive semidefinite.

Finally, the problem in (\ref{p2prime}) that is solved at the $\kappa$th iteration of MM can be rewritten as the following optimization:
%\begin{center}
\begin{alignat}{3}\label{pzegon}
  & \max_{t,\{\bx_{i}\}_{{i=1}}^N}   t &&&& \\
&\text{s.t.} &&\!\!\!\!\!\!C_i^{(\kappa-1)}+2\Re\{(\bb_i^{(\kappa-1)})^H\bx_i\}+&&\sum_{j=1}^{N}\bx_j^H\bG_{ji}^{(\kappa-1)}\bx_j\leq -t \nonumber\\
& && &&\forall i=1,2,\dots , N\nonumber\\
 & && \!\!\!\!\!\!\| \bx_{i} \|_{2}^2  \leq  p_{i} && \forall i=1,2,\dots , N \nonumber
\end{alignat}
where $C_i$ is the following real-valued constant:
\begin{align}\label{ci}
    C_i^{(\kappa-1)}=&-\log \det (\bU^H ({\bB}_{i}^{(\kappa-1)})^{-1}\bU)-\tr\{\bF_i^{(\kappa-1)}{\bB}_i^{(\kappa-1)}\}\\ \nonumber
    &+\tr\{(\bF_i^{(\kappa-1)})_{11}\}+\tr\{ (\bF_i^{(\kappa-1)})_{22}\bGamma_i \}.
  \end{align}

Note that (\ref{pzegon}) is a convex problem with a linear objective and quadratic constrains. Hence it can be expressed as a second-order cone program (SOCP). The proposed algorithm, which is based on iteratively solving \eqref{pzegon}, is summarized in Table II. In the first step, we initialize the algorithm with i.i.d. CSCG random variables after making them feasible by normalization i.e. $\| \bx_{i} \|_{2}^2  \leq  p_{i}$. In the second step, we use  efficient methods such as interior point algorithms to solve the problem in \eqref{pzegon} \cite{boyd2004convex}\cite{ben2011lectures}. The (conservative) computational complexity for this step is ${\cal{O}}(N^{4.5} M^6)$ assuming $M_i=M, \; \forall i$ \cite{ben2011lectures}. Step 3 includes matrix manipulations for updating the parameters; concretely, it includes inversion of the matrices $\{\bB_i \in \realS^{++}_{M_i+L_i}\}_{i=1}^N$ for obtaining $\bF_i$ via \eqref{F}   and then updating the parameters according to \eqref{gb}, \eqref{gb1}, and \eqref{ci}. The computational complexity of this step is mainly dominated by computing $\bB_i^{-1}$ which is of ${\cal O}((M_i+L_i)^3)$ for each $i$. After step 3, the stop criterion is checked and  steps 1 to 3 are repeated until this criterion is satisfied.

\rema{\normalfont (Calculating $\bV_{i}$ from $\bQ_{i}$):  At the convergence of the proposed method, the optimized transmit covariances $\{\bQ_{i}=\widetilde{\bV}_i \widetilde{\bV}_i^H\}\in \complexC^{M_i\times M_i}$ are obtained. Next, the precoder matrices $\{\bV_{i}\}\in \complexC^{M_i\times d_i}$ are obtained as  square roots of the $\{\bQ_{i}\}$: $\bV_{i}\bV_{i}^H=\bQ_{i}$. Note that the so-obtained precoder matrix $\bV_{i}$ is not unique but this has no effect on the rate. Indeed the rate $R_i$ in \eqref{Rate} is a many-to-one function of $\bV_{i}$ as $\bV_{i}$ and $\bV_{i}\bA$ lead to the same $R_i$ for any matrix $\bA$ satisfying $\bA\bA^H=\bI$. Also note that whenever $\bQ_{i}$ is (nearly) singular, one can perform a thresholding operation on its eigenvalues and reduce the number of columns of $\bV_{i}$ accordingly. Finally observe that the optimized stream lengths $\{d_{i}\}_{i=1}^N$ are given once we have $\{\bV_{i}\}\in \complexC^{M_i\times d_i}$.} \hfill $\blacksquare$\label{Rem1}
\normalfont

\subsection{Design of precoder matrices $\{\bV_i\}_{i=1}^{N}$ for given $\{d_i\}_{i=1}^{N}$}
In the previous subsection, we proposed an efficient algorithm to design the precoder covariance matrices $\{\bQ_i\}_{i=1}^{N}$. By using this algorithm, the optimum precoder matrices $\{\bV_i\}_{i=1}^{N}$ as well as the optimum number of their columns $\{d_i\}_{i=1}^{N}$  are designed. In some cases, the symbol stream length $d_i$ is given and  the precoder matrices  $\{\bV_i\}_{i=1}^{N}$ are the only ones to be designed. In such a case, the following optimization problem is considered:
\begin{align}\label{pv}
   \max_{\{\bV_{i}\}_{{i=1}}^N, t}& t  \\
\text{s.t.} \qquad &R_i  \geq  t  &   \forall i=1,2,\dots , N \nonumber   \\
 & \|\bV_i\|_F^2 \, \leq \,  p_{i}&   \forall i=1,2,\dots , N\nonumber
\end{align}
where $R_i$ is given in \eqref{Rate}. This problem, which is also NP-hard,  can be tackled by modifying the proposed method as follows. The matrices $\bU$ and $\bB_i$ defined in \eqref{U} and \eqref{bi} are replaced by:
\begin{equation}\label{Ubar}
  \overline{\bU}\triangleq\left[
                                                          \begin{array}{cc}
                                                            \bI_{d_i} & \textbf{0}_{d_i\times L_i} \\
                                                          \end{array}
                                                        \right]^T
\end{equation}
and
\begin{equation}\label{bibar}
\overline{\bB}_{i}=\left[
        \begin{array}{cc}
          \bI_{d_i} & \bV_{i}^{H}\bH_{ii}^{H} \\
          \bH_{ii}\bV_{i} & \bGamma_i+\displaystyle\sum_{j=1}^{N}\bH_{ji}\bV_{j}\bV_{j}^H\bH_{ji}^{H} \\
        \end{array}
        \right] ,
\end{equation}
respectively. By following an approach similar to that used in designing  $\{\bQ_i\}_{i=1}^{N}$, the following problem is solved at the $\kappa $th iteration of the algorithm to obtain $\{\bV_i\}_{i=1}^{N}$:
\begin{alignat}{3}\label{pzegonbar}
  & \max_{t,\{\overline{\bx}_{i}\}_{{i=1}}^N,}   t &&&& \\
&\text{s.t.} && \overline{C}_i^{(\kappa-1)}+2\Re\{(\overline{\bb}_i^{(\kappa-1)})^H\overline{\bx}_i\}+
&&\sum_{j=1}^{N}\overline{\bx}_j^H(\overline{\bG}_{ji}^{\kappa-1})^H\overline{\bx}_j\leq -t \nonumber\\
& && && \forall i=1,2,\dots , N\nonumber\\
 & && \| \overline{\bx}_{i} \|_{2}^2  \leq  p_{i} && \forall i=1,2,\dots , N \nonumber
\end{alignat}
where $\overline{\bx}_i\triangleq\vect(\bV_i$), $\overline{\bG}_{ji}\triangleq\bI_{d_i}\otimes(\bH_{ji}^H(\bF_i)_{22}\bH_{ji})$, and $\overline{C}_i^{(\kappa-1)}$ is the following real constant:
\begin{align}\label{cibar}
    \overline{C}_i^{(\kappa-1)}=&-\log \det (\overline{\bU}^H(\overline{{\bB}}_{i}^{(\kappa-1)})^{-1}\bU)-
    \tr\{\overline{\bF}_i^{(\kappa-1)}\overline{{\bB}}_{i}^{(\kappa-1)}\}\\
    &+\tr\{(\overline{\bF}_i^{(\kappa-1)})_{11}\}+\tr\{ (\overline{\bF}_i^{(\kappa-1)})_{22}\bGamma_i \}.\nonumber
  \end{align}
The problem in (\ref{pzegonbar}) is also a convex SOCP. Therefore $\{\bV_i\}_{i=1}^{N}$ can be obtained by slightly modifying the procedure in Table II according to the discussion above.

%\rema{ \normalfont  (Convergence): The proposed algorithm  can be shown to be locally convergent. To this end, observe that for the minimum rate at the $\kappa$th iteration, we have that:
%\begin{align}\label{Conv}
% & \min_{i}\, \log \det (\bU^{H}{(\bB_{i}^{(\kappa-1)})}^{-1}\bU)= \min_{i}\,
%   g_i^{(\kappa)}(\widetilde{\bV}^{(\kappa-1)}_{1},\cdots,\widetilde{\bV}^{(\kappa-1)}_{N})\\ \nonumber
% &  \leq \min_{i}\, g_i^\kappa(\widetilde{\bV}^{(\kappa)}_{1},\cdots,\widetilde{\bV}^{(\kappa)}_{N}) \leq
%   \min_{i}\, \log \det (\bU^{H}{(\bB_{i}^{(\kappa)})}^{-1}\bU) \nonumber
%\end{align}
% The first inequality in \eqref{Conv} holds due to the maximization step at the $\kappa$th iteration and the second one is satisfied due the definition of the
%minorizer, see \eqref{mnz}. Combining (\ref{Conv}) and the fact that  the objective function is upper bounded, it follows that the sequence of objective values converges.}
%
%\hfill $\blacksquare$

\subsection{Convergence}
In this subsection, we study the convergence of the proposed method and  prove that it converges to a stationary point of the problem. To this end, observe that for the minimum rate at the $\kappa$th iteration, we have that:
\begin{align}\label{Conv}
 & \min_{i}\,\! \log \! \det (\bU^{H}\!{(\bB_{i}^{(\kappa-1)})}\!^{-1}\!\bU)\!\!=\!\! \min_{i}\,\!
   g_i^{(\kappa)}(\widetilde{\bV}^{(\kappa-1)}_{1}\!,\!\cdots\!,\!\widetilde{\bV}^{(\kappa-1)}_{N})\\ \nonumber
 &  \leq \min_{i}\, g_i^{(\kappa)}(\widetilde{\bV}^{(\kappa)}_{1},\cdots,\widetilde{\bV}^{(\kappa)}_{N}) \leq
   \min_{i}\, \log \det (\bU^{H}{(\bB_{i}^{(\kappa)})}^{-1}\bU) \nonumber
\end{align}
 The first inequality in \eqref{Conv} holds due to the maximization step at the $\kappa$th iteration and the second one is satisfied due the definition of the
minorizer, see \eqref{mnz}. Combining (\ref{Conv}) and the fact that  the objective function is upper bounded, it follows that the sequence of objective values converges to a limit point $f^{\star}$. In the sequel, we prove that $f^{\star}$ is a stationary value and that the associated sequence $\{\bQ_i^{(\kappa)}\}$ converges to a stationary point of the design problem. The following theorem (from \cite{sun2017majorization}--see also \cite{razaviyayn2013unified}) establishes a general framework for convergence analysis of MM algorithms.

\thm{\normalfont  Consider the following optimization problem
\begin{align}\label{pc}
\!\max_{\bx}\, \quad & {f(\bx)} \\ \nonumber
 \text{s.t.} \qquad & \bx\in C
\end{align}
with $C$ being a compact convex set in $\realR^{N}$ and $f(\bx)$ being a non-smooth function (like our criterion in the design problem \eqref{p}). Let  $g(\bx,\bx_{0})$ be a minorizer of $f(\bx)$ at $\bx_{0}$. The sequence generated by MM algorithm for the problem \eqref{pc}  converges to a stationary point if the following conditions are satisfied:
\newline
\textbf{(A.1)} $g(\bx,\bx_{0})$ be continuous in $\bx$ and $\bx_{0}$.
\newline
\textbf{(A.2)} The sublevel set defined as $ lev_{\leq f (\bx_0 )}f := \{\bx \in C|f (\bx)
\leq f (\bx_{0} )\}$ is compact (given $f (\bx_{0}   ) \textless \infty$).
\newline
\textbf{(A.3)}
\begin{align}\label{A3}
  & \limsup_{\lambda\to0}\frac{g(\bx_{0}+\lambda \bd,\bx_{0})-g(\bx_{0},\bx_{0})}{\lambda}= \\ \nonumber
  & \limsup_{\lambda\to0}\frac{f(\bx_0+\lambda \bd)-f(\bx_0)}{\lambda}, \;\; \forall \bx_0+\bd \in C\nonumber
\end{align}
\hfill $\blacksquare$

Next, we prove the following theorem that lays the ground for convergence analysis of a class of maxmin optimization  problems (including the design problem in (\ref{p})) tackled by MM algorithms.
 \thm{\normalfont  Consider the following maxmin optimization problem
 \begin{align}\label{p2}
&\!\max_{\bx}\, \min_{i=1,2,\dots , M} \quad  {f_{i}(\bx)} \\ \nonumber
 &\text{s.t.} \qquad  \bx \in C
\end{align}
where $f_i(\bx), i=1,...,M$ are convex functions and $C$ is a compact convex set in  $\realR^{N}$.
Let $g_{i}(\bx,\bx_{0})\triangleq f_{i}(\bx_{0})+ \tr\{\left(\nabla_{\bx} f_{i}(\bx_{0})\right)^T(\bx-\bx_{0})\}$ be the minorizer of $f_i(\bx)$ at $\bx_0$ used in the MM method. Then, the sequence generated by the MM algorithm for the problem \eqref{p2} converges to a stationary point.
   }
\begin{proof}
{See Appendix F.}
\end{proof}

}

\lemm{
\normalfont The proposed MM method for the design problem \eqref{p} converges to a stationary point.

 \IEEEproof {To prove this lemma, we follow the proof provided  for Theorem 2 in Appendix F.  First, note that every $ R_{i}$ is continuous and their minimizers are continuous. Also, the  constraint set in  problem \eqref{p}  is closed, convex, and bounded in the matrix space with Frobenius norm. Thus the conditions \textbf{A.1} and \textbf{A.2} are satisfied. Second,  every $ R_{i}$  is convex with respect to $ \bB_{i}$. Therefore, according to Appendix F, we can write:
\begin{align}\label{L21}
&\limsup_{\lambda\to0}\frac{g(\bar{\bB_{i}}+\lambda \bD,\bar{\bB_{i}})-g(\bar{\bB_{i}},\bar{\bB_{i}})}{\lambda}=\\ \nonumber &\limsup_{\lambda\to0}\frac{f(\bar{\bB_{i}}+\lambda \bD)-f(\bar{\bB_{i}})}{\lambda}
\end{align}
To complete the proof, we use the chain rule to show that \textbf{A.3} is also satisfied for our problem \cite{yang1998generalized}:
%\begin{align} \nonumber
%&\limsup_{\lambda\to0}\frac{g({\widetilde{\bV}_i}+\lambda \bD,{\widetilde{\bV}_i})-g(\widetilde{\bV}_i,{\widetilde{\bV}_i})}{\lambda}= \nonumber \\ \nonumber
%&\limsup_{\lambda\to0}\frac{f(\widetilde{\bV}_i+\lambda \bD)-f(\widetilde{\bV}_i)}{\lambda}
%\end{align} \hfill $\blacksquare$}
\begin{align} \nonumber
&\limsup_{\lambda\to0}\frac{g({\bQ_i}+\lambda \bD,{\bQ_i})-g(\bQ_i,{\bQ_i})}{\lambda}= \nonumber \\ \nonumber
&\limsup_{\lambda\to0}\frac{f(\bQ_i+\lambda \bD)-f(\bQ_i)}{\lambda}
\end{align} } \hfill $\blacksquare$}}

%
%\rema{  \normalfont (Precoder design for given $\{d_{i}\}_{i=1}^N$):  As explained above, by designing the precoder covariance matrices $\{\bQ_i\}_{i=1}^{N}$, we simultaneously design the  precoder matrices $\{\bV_i\}_{i=1}^{N}$ and the  number of their columns $\{d_i\}_{i=1}^{N}$, i.e. the lengths of the symbol streams. In some cases, the length of  symbols streams $\{d_i\}_{i=1}^N$ are given and the precoder matrices  $\{\bV_i\}_{i=1}^{N}$ should be designed directly. To deal with this case, we can modify the proposed method simply by  replacing $\widetilde{\bV}_i\in \complexC^{M_i\times M_i}$ in \eqref{pmin} with $\bV_i\in \complexC^{M_i\times d_i}$.\hfill $\blacksquare$%  and following a procedure  similar to that for $\{\bQ_i\}_{i=1}^{N}$. }

\begin{table}[tp]
\footnotesize
\caption{The proposed method  for the max-min rate design of the transmit covariance matrices in MIMO-IC.} \label{table:method} \centering
\begin{tabular}{p{3.3in}}
\hline \hline
%\\
\textbf{Step 1}: Initialize $\{\bx_i\}_{i=1}^{N}$ with complex random  vectors in $\complexC^{M_i^2\times 1}$ such that they satisfy $\| \bx_{i} \|_{2}^2  \leq  p_{i}$.\\
\textbf{Step 2}: Solve the (convex) SOCP problem in (\ref{pzegon}).\\
\textbf{Step 3}: Update $\bb_i$,  $\bG_{ji}$, and $C_i$ according to equations \eqref{F}, \eqref{gb}, \eqref{gb1}, and \eqref{ci}, respectively. \\
\textbf{Step 4}: Repeat steps 1 and 2 until a pre-defined stop criterion is satisfied,
e.g $|t^{(\kappa)}-t^{(\kappa-1)}|\leq\epsilon$, for a given $\epsilon >0$.\\
\hline \hline
\end{tabular}
\end{table}
\normalfont

\normalfont
\section{Precoder Design in the presence of a priori knowledge uncertainty}
In practice there always exist uncertainties in the noise covariance and the channel state information. In this section we will consider these uncertainties in the design problem.

 We first consider the effect of imperfect CSI due to  channel estimation errors.  Using the conventional LMMSE estimator, the  channels can be modeled as \cite{kay1993fundamentals}:
 \begin{equation}\label{Est}
   \bH_{ji} = \hat{\bH}_{ji} + \bZ_{ji}
 \end{equation}
 where $\hat{\bH}_{ji}$ is  the estimate of the true channel ${\bH}_{ji}$ and $\bZ_{ji}$ is the channel estimation error which is assumed to be uncorrelated  with $\hat{\bH}_{ji}$. Assuming the entries of $\bH_{ji}$ are i.i.d random variables (RVs) with variances $\sigma^2_{ji}$, the entries of $\hat{\bH}_{ji}$  and $\bZ_{ji}$ will be i.i.d RVs with variances $\rho_{ji}^2 \sigma^2_{ji}$ and $(1-\rho_{ji}^2 )\sigma^2_{ji}$, respectively. The parameter $\rho_{ji}\in [0,1]$ quantifies the estimation accuracy, in particular if $\rho_{ji}=1$, $\hat{\bH}_{ji} = {\bH}_{ji}$ and CSI is perfect.% can be calculated from training sequence, receiver noise and channel characteristics. For case where orthogonal training sequences with length $N_t$ and power $p_{t}$ are employed for channel estimation, $\rho_{ji}^2=\frac{1}{1+\frac{\sigma^2_n}{p_{t}N_t\sigma^2_{ji}}}$, where $\sigma^2_n$ is the variance of the receiver noise. }.

 Substituting \eqref{Est} in \eqref{model}, we obtain:
 \begin{equation}\label{model2}
  \by_i=\underbrace{\hat{\bH}_{ii}{\bV'}_i\bs_i}_{\textrm{desired signal}}+\underbrace{\sum_{j\neq i}\hat{\bH}_{ji}{\bV'}_j\bs_j+
  \sum_{j=1}^N{\bZ}_{ji}{\bV'}_j\bs_j+\bn_i}_{\textrm{interference plus estimation error and noise}}
\end{equation}
with  ${\bV'}_i$ being the precoder matrix of the $i$th transmitter designed under imperfect CSI. It can  be proved that:
\begin{equation}\label{Exp}
  \mathbb{E}\left\{{\bZ}_{ji}{\bV'}_j{\bV'}_j^H{\bZ}_{ji}^H\right\}= (1-\rho_{ji}^2 )\sigma^2_{ji}\tr\{{\bV'}_j{\bV'}_j^H\}\bI_{L_i}
\end{equation}
Therefore, the LMMSE decoder will be:
\begin{align}\label{LMMSE2}
  \hat{\bW}_i^{\textrm{LMMSE}}=&{\bV'}_i^H \hat{\bH}_{ii}^H\left(\sum_{j=1}^N \hat{\bH}_{ji} {\bV'}_j {\bV'}_j^H \hat{\bH}_{ji}^H +\right. \\ \nonumber
  & \left. \sum_{j=1}^N (1-\rho_{ji}^2 )\sigma^2_{ji}\tr\{{\bV'}_j{\bV'}_j^H\}\bI_{L_i} + \bGamma_i \right)^{-1}
\end{align}
Let ${\bQ'}_j\triangleq{\bV'}_j{\bV'}_j^H$ be the precoder covariance matrices in the imperfect CSI case. Note that the term $\sum_{j=1}^N{\bZ}_{ji}{\bV'}_j\bs_j$ in (\ref{model2}) is the sum of the products of Gaussian random variables (i.e. $\bZ_{ji}$ and $\bs_j$) and hence, it is no longer Gaussian; this observation leads to difficulties for computation of the user rate. Therefore, in this case, we resort to a common approach in the literature (see e.g. \cite{wang2012efficient,ngo2013energy,ho2011decentralized} and references therein) to make the problem tractable; more precisely, the following lower bound $\hat{R}_i$ on the rate of the $i$th user is considered as the design metric:
 %The rate\footnote{Note that according to the Jensen's inequality, \eqref{RateEst}  gives a lower bound (approximation) on the transmission rate.\footnote{The tightness of the bound has been investigated in \cite{yoo2006capacity} and it has been shown that it is usually tight for Gaussian transmit signal.}
%This approximation has been widely used in the literature to make the transmission rate tractable in the presence of imperfect CSI (e.g. see  \cite{wang2012efficient,ngo2013energy,ho2011decentralized} and references therein).} $\hat{R}_i$ of the $i$th user for this case becomes:
 \begin{align}\label{RateEst}
  {\hat{R}}_i  =  \log \det \! & \left(\bI_{L_i}+\hat{\bH}_{ii} {\bQ'}_i \hat{\bH}_{ii}^{H}\! \left[\bGamma_i+ \sum_{j \neq i}^{} \hat{\bH}_{ji} {\bQ'}_j \hat{\bH}_{ji}^H \right. \right. \\ \nonumber
    & \left. \left. +\sum_{j=1}^N (1-\rho_{ji}^2 )\sigma^2_{ji}\tr\{\bQ'_j\}\bI_{L_j}\right]^{-1}\right) \nonumber
\end{align}

%Finally, the max-min problem under imperfect CSI assumption can be cast:
%\begin{align}\label{pEst}
%\!\max_{\{{\bQ'}_{i}\}_{{i=1}}^N}\, &\min_{i=1,2,\dots , N} \quad  {{R'}_i} \\ \nonumber%
% \text{s.t.} \quad & \tr\{{\bQ'}_i\} \, \leq \, p_{i}, \qquad & \forall i=1,2,\dots , N \\ \nonumber%
% &{\bQ'}_i\succeq\textbf{0}  \quad & \forall i=1,2,\cdots, N \nonumber
%\end{align}
%with ${R'}_i$ given in \eqref{RateEst}. This problem is also NP-hard and can efficiently be dealt with by modifying the proposed method for perfect CSI.  We can obtain the solutions to the problem \eqref{pEst} using the iterative algorithm given in Table II by substituting $\bGamma_i$ in \eqref{bi} with $\hat{\bGamma}_i=\bGamma_i+\sum_{j=1}^{N}(1-\rho_{ji}^2 )\sigma^2_{ji}\tr\{\hat{\bQ}_j\}\bI_L$ and consequently modifying $\bF_i$, $\bG_{ji}$ and $C_i$.

Next, we also consider the uncertainty of the noise covariance matrices, which can be modeled as \cite{naghsh}:
\begin{equation}\label{uncer}
  \|\bGamma_i - \hat{\bGamma}_i\|_2 \leq \zeta_i, \quad \forall i=1,\cdots, N
\end{equation}
where $\hat{\bGamma}_i$s are known positive definite matrices (initial guesses of the covariance matrices) and $\zeta_i$s are positive scalars that determine the size of the uncertainty regions.
 We  remark on the fact that in the case of imperfect CSI, we have uncertainty about the true channel value; however, in the case of noise vectors, we consider uncertainty in the noise \emph{covariance} matrix (rather than in the noise vector). Therefore, we consider a worst-case approach to deal with  uncertain  noise covariance matrices which is commonly used in literature (see for instance \cite{karbasi2015knowledge,dong2006finite}).

We can robustify the design method with respect to a priori knowledge uncertainty by considering the following reformulation of the optimization problem:
\begin{alignat}{3} \label{pun}
&\!\max_{\{{\bQ'}_{i}\}_{{i=1}}^N}\,&& \min_{i=1,\dots , N} \min_{\{{\bGamma_i}\}_{{i=1}}^N} \quad  {\hat{R}_i} && \\ \nonumber
&\text{s.t.} \qquad &&  \tr\{{\bQ'_i}\} \, \leq \, p_{i}, \quad && \forall i=1,2,\dots , N \\\nonumber%
& && \|\bGamma_i - \hat{\bGamma}_i\|_2 \leq \zeta_i, \quad && \forall i=1,2,\cdots, N \\\nonumber
& &&\bGamma_i\succeq\textbf{0}, {\bQ'_i}\succeq\textbf{0}  \quad && \forall i=1,2,\cdots, N \nonumber
\end{alignat}
where ${\hat{R}}_i$ is as given in \eqref{RateEst}. In what follows we present a theorem which shows that the problem in \eqref{pun} can be dealt with via a modified version of the  method proposed in Section III.
\thm{
\normalfont Let $R'_i$ be defined as:
 \begin{align}\label{Rateun}
  R'_i  =  \log \det \! & \left(\bI_{L_i}+\hat{\bH}_{ii} {\bQ'}_i \hat{\bH}_{ii}^{H}\! \left[\bGamma'_i+ \sum_{j \neq i}^{} \hat{\bH}_{ji} {\bQ'}_j \hat{\bH}_{ji}^H \right. \right. \\ \nonumber
    & \left. \left. +\sum_{j=1}^N (1-\rho_{ji}^2 )\sigma^2_{ji}\tr\{\bQ'_j\}\bI_{L_j}\right]^{-1}\right) \nonumber
\end{align}
where
$\bGamma'_i=\hat{\bGamma}_i+ \zeta_i \bI_{L_i}$.
The problem
\begin{alignat}{3} \label{pun2}
&\!\max_{\{{\bQ'_i}\}_{{i=1}}^N}\,&& \min_{i=1,2,\dots , N} \quad  {R'_i} && \\ \nonumber
&\text{s.t.} \qquad &&  \tr\{{\bQ'_i}\} \, \leq \, p_{i}, \quad && \forall i=1,2,\dots , N \\\nonumber%
 &&&{\bQ'_i}\succeq\textbf{0}  \quad && \forall i=1,2,\cdots, N \nonumber
\end{alignat}
is equivalent to the problem in \eqref{pun} in the sense that these two problems share the same solution $\{{\bQ'_i}\}_{{i=1}}^N$.

 \IEEEproof Noting that the inner problem of \eqref{pun} is separable w.r.t $i$, we consider it for a fixed $i$:
 \begin{align} \label{pun3}
&\min_{{\bGamma_i}\succeq\textbf{0}} \quad  {\hat{R_i}}  \\ \nonumber
&\text{s.t.} \qquad   \|\bGamma_i - \hat{\bGamma}_i\|_2 \leq \zeta_i\quad&&  \nonumber%
\end{align}
Note that $\|\bGamma_i - \hat{\bGamma}_i\|_2=\sqrt{\lambda_{max}\left((\bGamma_i - \hat{\bGamma}_i)^H(\bGamma_i - \hat{\bGamma}_i)\right)}$ and the matrix $\bGamma_i - \hat{\bGamma}_i$ is Hermitian; therefore, the constraint $\|\bGamma_i - \hat{\bGamma}_i\|_2 \leq \zeta_i$ is equivalent to $\max_m |\lambda_{m}(\bGamma_i - \hat{\bGamma}_i)| \leq \zeta_i$ with $\lambda_m(\bGamma_i - \hat{\bGamma}_i)$ being the $m$th eigenvalue of the matrix $\bGamma_i - \hat{\bGamma}_i$. Therefore, we have that
\begin{equation}\label{lambda}
  \lambda_m(\bGamma_i - \hat{\bGamma}_i) \in [-\zeta_i,\zeta_i], \forall m=1,2,..., L_i
\end{equation}
 Consequently, it can be verified that the constraint in \eqref{pun3} is equivalent to
\begin{equation}\label{con1}
  \hat{\bGamma_i}-\zeta_i\bI_{L_i} \preceq \bGamma_i\preceq \hat{\bGamma_i}+\zeta_i\bI_{L_i}
\end{equation}
and therefore, that the problem \eqref{pun3}  is equivalent to the following optimization:
 \begin{align} \label{pun4}
&\min_{{\bGamma_i}\succeq\textbf{0}} \quad  {\hat{R}_i} \\ \nonumber
&\text{s.t.} \qquad \hat{\bGamma}_i-\zeta_i\bI_{L_i} \preceq \bGamma_i\preceq \hat{\bGamma}_i+\zeta_i\bI_{L_i} \\\nonumber%
\end{align}
Note that $ \bH_{ji} \bQ_j \bH_{ji}^H\succeq \textbf{0}$ and also that $(1-\rho_{ji}^2 )\sigma^2_{ji}\tr\{\bQ'_j\}\bI_{L_j}\succeq \textbf{0}, \forall i,j$. Consequently, using  \eqref{con1} we  have that:
 \begin{align}\label{con2}
  &\left[\bGamma_i+ \!\!\sum_{j \neq i}^{} \bH_{ji} \bQ_j \bH_{ji}^H+\!\!\sum_{j=1}^N (1-\rho_{ji}^2 )\sigma^2_{ji}\tr\{\bQ'_j\}\bI_{L_j}\!\!\right]^{-1} \!\!\!\!\! \succeq \quad \quad \\ \nonumber
  & \left[\hat{\bGamma}_i+ \zeta_i \bI_{L_i} + \!\!\sum_{j \neq i}^{} \bH_{ji} \bQ_j \bH_{ji}^H+\!\!\sum_{j=1}^N (1-\rho_{ji}^2 )\sigma^2_{ji}\tr\{\bQ'_j\}\bI_{L_j}\!\!\right]^{-1} \nonumber
\end{align}
%Considering \eqref{con2}, \eqref{RateEst} and \eqref{pun}, we conclude that $\hat{R}_i \geq R'_i$ with equality when ${ \bGamma_i^\star}=\bGamma'_i=\hat{\bGamma}_i+ \zeta_i \bI_{L_i}$, which completes the proof.
The stated result follows from  (\ref{con2}).
 \hfill $\blacksquare$}

 \cor{\normalfont The robust design problem in \eqref{pun} can be solved using the proposed algorithm (see Table II) after replacing $\bGamma_i$ with  $\bGamma'_i+\sum_{j=1}^{N}(1-\rho_{ji}^2 )\sigma^2_{ji}\tr\{{\bQ}^{\prime}_j\}\bI_{L_j}$,  and after modifying $\bF_i$, $\bG_{ji}$ and $C_i$ accordingly.}\hfill $\blacksquare$

 \normalfont

\section{Numerical Results}
In this section, we present several numerical examples to illustrate the performance of the proposed method.
In all cases, unless otherwise stated, we assume that $N=3$,  $M_i=L_i\triangleq M=4$,  and SNR$\triangleq \frac{L_i p_i}{\tr\{\bGamma_i \}}$=15dB, $\forall i=1,2,\cdots,N$. The receiver noise vectors are assumed to be white with unit variances, i.e., $\bGamma_i=\bI_{L_i}$, and the elements of channel matrices are i.i.d. CSCG random variables with zero mean and unit variances. The convex problems are solved using CVX toolbox. We set $\epsilon=10^{-3}$ for the stop criterion of the algorithm.
\begin{figure}[t]
\centering
\includegraphics[width= 8.5 cm, height =7 cm] {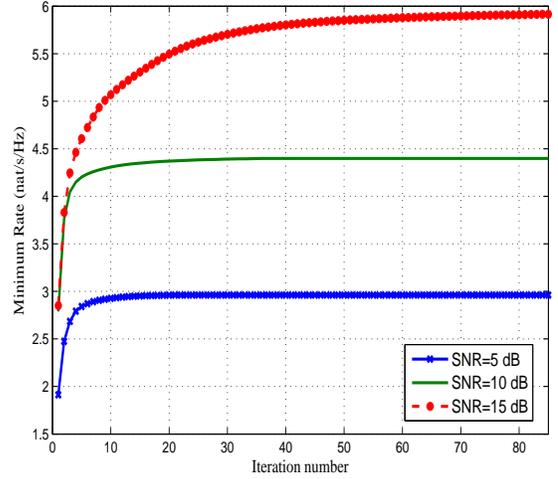} \\
\caption{The min rate versus number of iterations for the proposed method.}
\label{iter}
\end{figure}
%\begin{figure}[t]
%\centering
%\includegraphics[width= 8.5cm, height =7 cm] {DIF_sum_min} \\
%\caption{Comparison between min-rate maximization of the proposed method and the method in \cite{razaviyayn2013linear} for $N=10$.}% and sum-rate maximization for $N=10$. The rate of users 4 and 8 are too small to be visible. }
%\label{comp}
%\end{figure}
\begin{figure}[t]
\centering
\includegraphics[width= 8.5cm, height =7 cm] {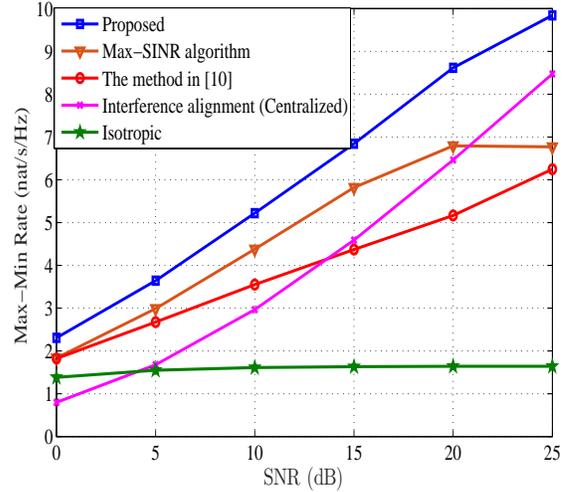} \\
\caption{The max-min rate (versus SNR) achieved by the proposed algorithm, the method in \cite{razaviyayn2013linear}, two methods in \cite{gomadam2008approaching}, and the isotropic transmission.}
\label{minratesnr}
\end{figure}
\begin{figure}[t]
\centering
\includegraphics[width= 8.5cm, height =7 cm] {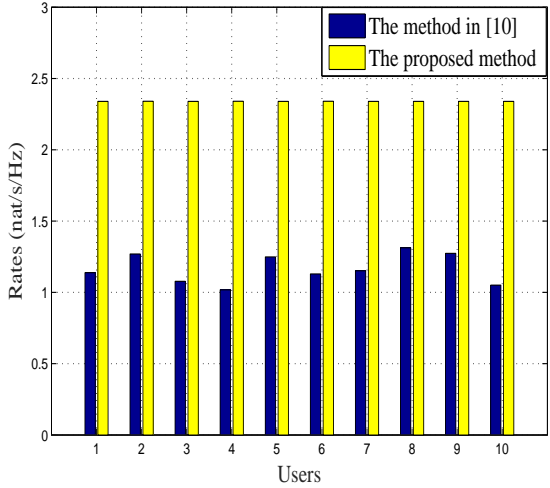} \\
\caption{Illustration of users' rates for min-rate maximization algorithms for $N=10$: the proposed method and the method in \cite{razaviyayn2013linear}.}% and sum-rate maximization for $N=10$. The rate of users 4 and 8 are too small to be visible. }
\label{comp}
\end{figure}

To investigate the convergence behaviour of the proposed algorithm, in \Fig\ref{iter} we plot the  minimum rate achieved at various iterations  for different values of the SNR. It can be observed that the minimum rate, i.e. the value of the objective function, increases at each iteration in agreement with the results  in Section III. As expected, the higher the SNR, the larger the minimum rate.

%
%Next, we compare the proposed method with the method in \cite{razaviyayn2013linear} that has been suggested for min rate optimization. \Fig\ref{comp} illustrates the rates of various users for $N=10$. It can be seen that by using our method the rate will be distributed more fairly among the users; and the min rate associated with the proposed method is larger than that of \cite{razaviyayn2013linear}.% the proposed method leads to As expected, when the sum-rate is maximized, there exist users with very low rates (e.g, users 4 and 8) and users with  high rates (e.g. users 1 and 2). Thus, the major benefit of our proposed method in comparison with sum rate maximization methods like that in \cite{razaviyayn2012linear} is that by using our method the rate will be distributed more fairly among the users.

Next, we compare the proposed method with the method in \cite{razaviyayn2013linear} that has been suggested for min rate optimization. We also include in this comparison the interference alignment and the max SINR algorithms proposed in \cite{gomadam2008approaching} along with the isotropic transmission (in which a precoder matrix $\bV_i$ is given by a scalar version of identity matrix)\footnote{Note that the method in \cite{razaviyayn2013linear} considers min-rate optimization for given $d_i$ (see III.B). On the other hand, the  interference alignment and the max SINR algorithms proposed in \cite{gomadam2008approaching} consider sum-rate maximization; however, the min-rate associated with these two methods can be computed by employing the designed precoders of these methods.} as benchmarks.  \Fig\ref{minratesnr} shows the max-min rates, averaged over 30 random channel realizations. In this example, we set $d_i=2, \forall i$ for all methods (see subsection III.B). We observe that the rate obtained by the proposed algorithm is considerably higher than  those obtained by the other methods. This observation  shows that the method introduced in this paper can provide higher quality solutions to the design problem, (\ref{p}), than its competitors.
 %This might be due to the special reformulation\footnote{We herein remark on the fact that the design problem in (\ref{p}) is non-convex; hence, the devised method and the method in \cite{razaviyayn2013linear} cannot provide the associated global solution in general. However, the way of dealing with this non-convex problem affects the quality of the obtained solutions (see e.g., Figs. \ref{minratesnr} and \ref{200init}). Indeed, in this paper, we applied MM technique to the objective and employed a tight linear minorizer to tackle the problem. On the other hand, in \cite{razaviyayn2013linear}, the rate of the $i$th user is reformulated using the mean squared error matrix $\bE_i$ of each user (depending on $\bV_i$ and $\bW_i$) as well as an auxiliary variable $\widetilde{\bW}_i$. Then, for given $\{d_i\}$, the design problem is tackled via an alternation between precoders $\{\bV_i\}$, decoders $\{\bW_i\}$ and auxiliary variables $\{\widetilde{\bW}_i\}$.} employed in \cite{razaviyayn2013linear}.
 As expected, we also see that the  rates improve as SNR increases. %,  while for the method of \cite{razaviyayn2013linear} the  rate exhibits a saturation beyond an SNR of about 15dB. This behaviour is also seen in the simulation results reported in \cite{razaviyayn2013linear}.
 To further compare the proposed method with that of \cite{razaviyayn2013linear} which deal with the same maxmin design problem (for given $d_i$), we show in  \Fig\ref{comp} the rates of various users for $N=10$. It can be seen that by using our method the rate will be distributed more fairly among the users; and the min rate associated with the proposed method is larger than that of \cite{razaviyayn2013linear}. The superiority of the proposed method may be due to the special reformulation of the problem. More precisely,
  the design problem  is non-convex; hence, our devised method and the method in \cite{razaviyayn2013linear} are not guaranteed to find the global solution in general. Consequently, the way of dealing with this non-convex problem affects the quality of the obtained solutions. In this paper, we apply an MM technique to the objective function and employ a tight linear minorizer to tackle the problem. On the other hand, in \cite{razaviyayn2013linear}, the design problem is tackled via a BCD method, viz. an alternation between precoders, decoders and some auxiliary variables.
   It is worth mentioning that at each iteration of the BCD, the updating  of the vector variable is performed only in the direction of one block. Such a limitation of the BCD algorithm usually
leads to a relatively poor performance  when compared to that of other optimization algorithms.
 To gain further insights into these  methods, we consider a SIMO interference channel  and compare  the following methods: the proposed method, the method in \cite{razaviyayn2013linear}, the SDP bisection algorithm (SDPBA) \cite{liu2013max} and the  inexact cyclic coordinate ascent algorithm (ICCAA) \cite{liu2013max}. Fig. \ref{SIMO} shows the max-min rates  versus the number of transmit-receive pairs for the aforementioned methods. It is seen that  the proposed method performs close to the \emph{optimal} method in \cite{liu2013max} which is a global solver for the SIMO design problem. Also, similar to Fig. \ref{minratesnr},  the proposed method outperforms the method in \cite{razaviyayn2013linear}.% This observation lays the ground for expecting better performance of the proposed method when compared to that in \cite{razaviyayn2013linear} in MIMO case.

To provide more insights,  we next compare the computational complexities of the proposed method and the method in \cite{razaviyayn2013linear}. To this end, in Fig. \ref{Time}, we plot the average run time (for 100 random trials) of these methods  using a standard PC (with CPU Core i7 and 16 GB of RAM).  The average run time of the method in \cite{razaviyayn2013linear} is lower than that of the proposed method. On the other hand, as shown e.g. in Fig. \ref{minratesnr}, the method of this paper significantly outperforms that of \cite{razaviyayn2013linear} in terms of the minimum rate. %in cost of higher computational burden; this is a tradeoff between the complexity and the performance.

\begin{figure}[t]
\centering
\includegraphics[width= 8.5 cm, height =7 cm] {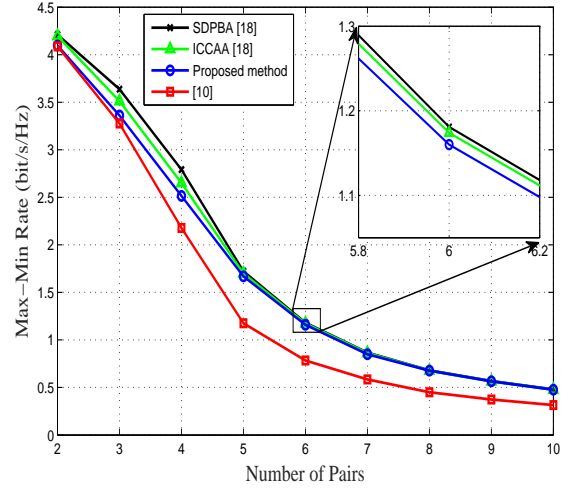} \\
\caption{Max-min rate of different methods versus the number of transmit-receive pairs in a SIMO scenario.}
\label{SIMO}
\end{figure}

\begin{figure}[t]
\centering
\includegraphics[width= 9.5 cm, height =7 cm] {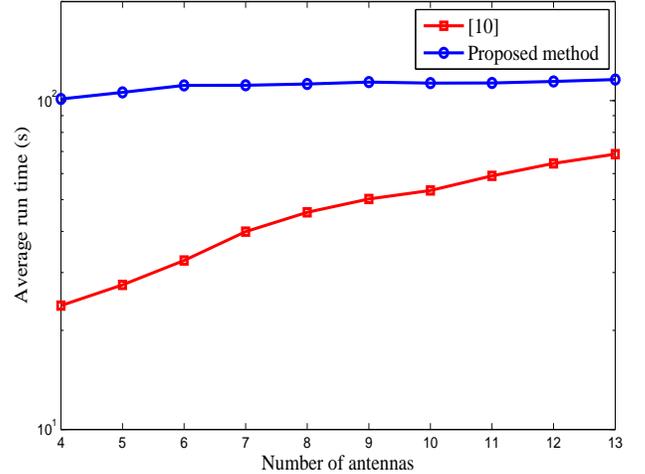} \\
\caption{Average run times of the proposed method and the algorithm in [10], versus the number of antennas.}
\label{Time}
\end{figure}

\begin{figure}[t]
\centering
\includegraphics[width= 8.5cm, height =7 cm] {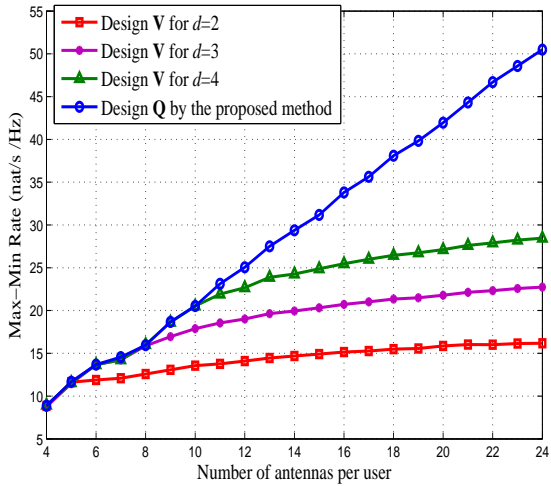} \\
\caption{The max-min rate achieved by the proposed algorithm  for designing $\{\bQ_i\}_{i=1}^{N}$ and, respectively, designing $\{\bV_i\}_{i=1}^{N}$ for different values of $d$,  versus the number of antennas per user ($M$).}
\label{QV}
\end{figure}

\begin{table}
\centering
\caption{Average run times for covariance design and for precoder design cases using the proposed method.} \label{table:times}
 \begin{tabular}{|c | c | c | c | c | c |}
 \hline
 \!\!\! \!\!\! & $\bQ_i$ design &
  \begin{tabular}{c}
    $\bV_i$ design  \\
    ($d=2$) \\
  \end{tabular}  &
    \begin{tabular}{c}
    $\bV_i$ design  \\
    ($d=3$) \\
  \end{tabular}  &
    \begin{tabular}{c}
    $\bV_i$ design  \\
    ($d=4$) \\
  \end{tabular}   \\ \hline
 $M=4$ & $98.8$s & $74.3$s & $88.3$s & $100.5$s   \\ \hline
 $M=6$ &$121.7$s  & $95.6$s & $108.4$s & $110.9$s \\
 \hline
 \end{tabular}
\end{table}

\begin{figure}
    \centering
    \begin{subfigure}
        \centering
        \includegraphics [width= 8.5cm, height =7 cm] {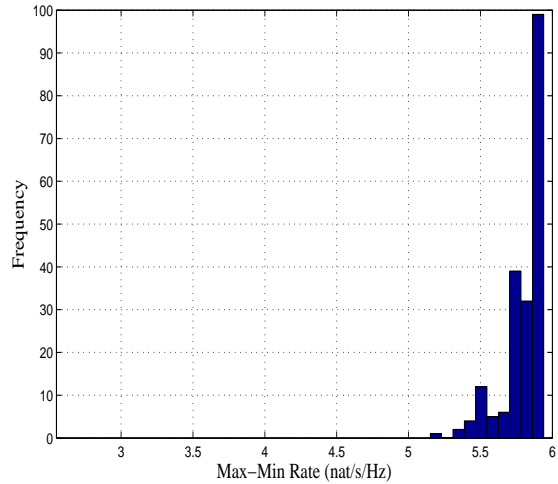}

    \end{subfigure}(a)
    \begin{subfigure}
        \centering
        \includegraphics [width= 8.5cm, height =7 cm]{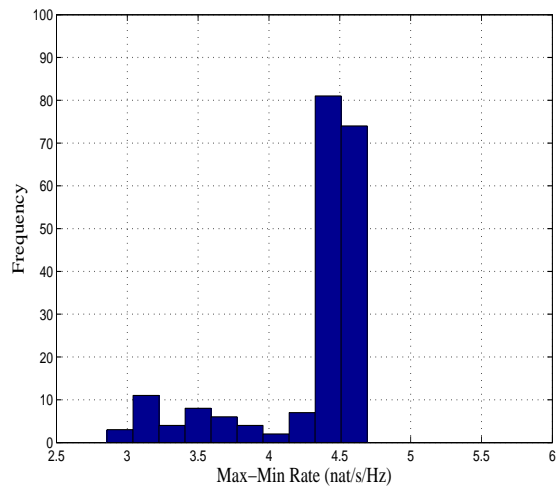}
    \end{subfigure}(b)

    \caption{Histogram of  max-min rates achieved using (a) the proposed method and (b) the  algorithm in\cite{razaviyayn2013linear} for 200 randomly chosen initial points.}\label{200init}
\end{figure}

Unlike the method in \cite{razaviyayn2013linear} that directly designs the precoder matrices  $\{\bV_i\}_{i=1}^{N}$ (given $\{d_i\}_{i=1}^{N}$), the proposed algorithm designs the precoder covariance matrices $\{\bQ_i\}_{i=1}^{N}$  (the precoder matrices $\{\bV_i\}_{i=1}^{N}$ can be obtained as a by-product of the proposed method). Therefore, by using the proposed method, the optimum precoder matrices $\{\bV_i\}_{i=1}^{N}$ as well as the optimum number of their columns $\{d_i\}_{i=1}^{N}$ (i.e. the lengths of  symbol streams) will be determined. To show the importance of this design aspect,  in \Fig\ref{QV} we plot the max-min rate achieved by  designing $\{\bQ_i\}_{i=1}^{N}$ and, respectively, by designing $\{\bV_i\}_{i=1}^{N}$ for certain values of  $\{d_i\}_{i=1}^{N}=d, \forall i$, versus the number of antennas.
As expected, the  rates achieved by designing $\{\bQ_i\}_{i=1}^{N}$ are higher  than (or equal to) those obtained by designing $\{\bV_i\}_{i=1}^{N}$ with fixed $\{d_i\}_{i=1}^{N}$. This can be explained by the fact that the optimal values for $\{d_i\}_{i=1}^{N}$ are also determined in the design of $\{\bQ_i\}_{i=1}^{N}$.
To compare the computational time for covariance design with that for only precoder design, we report the corresponding average run times of these two cases in  Table III for 100 random trials. We observe that there is no considerable difference between the average run times.

As stated earlier, the considered optimization problem is NP-hard and, as a result, any solution depends on the employed initial point. To investigate the dependency of the proposed method on the employed initial points,  in \Fig\ref{200init}.a we plot  the histogram of the max-min rates corresponding to 200 randomly chosen initial points. The histogram for the algorithm in \cite{razaviyayn2013linear} is also depicted in \Fig\ref{200init}.b. The  rates achieved by the proposed method are in the interval $[5.11-5.98]$ with a variance of about 0.02, while those achieved by the method in \cite{razaviyayn2013linear} are in the interval $[2.77-4.78]$ with a variance of about 0.2. Consequently, in this example, the proposed method achieves  higher  rates and its performance depends on the  initial points only mildly.

%of these two cases which supports preferability of covariance matrix design.

\begin{figure}[t]
\centering
\includegraphics[width= 8.5cm, height =7 cm] {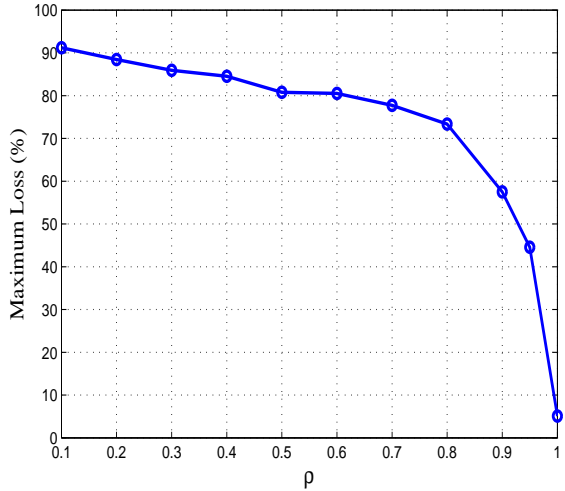} \\
\caption{The Loss ${\cal L}$ (in percentage)   versus the CSI error parameter $\rho$ ($\zeta=0.25$).}
\label{CSI}
\end{figure}
%\begin{figure}[t]
%\centering
%\includegraphics[width= 8.5cm, height =7 cm] {zetasmooth} \\
%\caption{The max-min rate achieved by the robust method proposed in Section IV versus uncertainty parameter $\zeta$.}
%\label{zeta}
%\end{figure}

Finally, we study the effect of channel estimation errors and noise covariance uncertainty on the performance of the  method proposed in Section IV. To this end, we set $\rho_{ji}=\rho$ as well as $\zeta_i=\zeta, \; \forall i,j=1,\cdots, N$ and define the loss parameter
 \begin{equation}\label{Loss}
   {\cal{L}}(\rho,\zeta)=1-  \frac{R_{nr}(\rho,\zeta)}{R_r(\rho,\zeta)}
 \end{equation}
 where $R_{nr}(\rho,\zeta)$ and $R_r(\rho,\zeta)$ denote the max-min rates achieved by the non-robust and the robust methods, respectively, for uncertainty parameters $(\rho,\zeta)$. Note that the loss parameter ${\cal{L}}(\rho, \zeta)$ quantifies the performance degradation caused by  employing the non-robust method instead of the robust one. Note also that ${\cal L}$ depends on the realizations of the channel matrices as well as noise covariances. In Fig. \ref{CSI}, we plot the maximum value of ${\cal{L}}(\rho, \zeta)$ versus $\rho$ for  $100$ realizations of channel matrices. In this example,  we set $\bGamma_i=\hat{\bGamma}_i+\zeta \bI, \forall i$ with $\zeta=0.25$. It can be seen that even for large values of $\rho$ (i.e., relatively low channel estimation errors), employing the robust method provides a significantly larger max-min rates. As expected,
  the loss decreases as the estimation quality improves, i.e., as $\rho$ increases. Note  that  the loss is non-zero even for  the case of $\rho=1$ in which CSI is perfect. This is due to the uncertainty in the noise covariances. Finally, note that in this example we have numerically observed that the performance loss is more sensitive to CSI uncertainty than the noise covariance uncertainty.

\section{Conclusion}
In this paper, we considered a MIMO  interference channel network with conventional LMMSE decoder matrices at the receivers for which we designed the transmit covariance matrices under the max-min fairness criterion. The problem is non-convex and NP-hard in the number of users. We proposed an efficient algorithm based on the MM optimization technique that provides quality  solutions to this design problem. We proved that the proposed algorithm converges to stationary points of the problem. Our results on the convergence analysis can pave the way for convergence analysis of other MM solvers for a class of maxmin optimization problems. We also considered  uncertainties in the  noise covariances and the CSI, and extended our algorithm to design precoder covariance matrices in these cases. Numerical results were included to illustrate the effectiveness of the proposed method in various scenarios.

\appendices

\section{Proof of \eqref{LMMSE} and \eqref{Rate} }\label{Appendix A}
We begin by proving the expression of the LMMSE in (\ref{LMMSE}). Assuming $\mathbb{E}\{\by_i\}=\textbf{0}$ and $\mathbb{E}\{\bs_i\}=\textbf{0}$, the LMMSE estimator of $\bs_i$ for given $\by_i$ has the following expression
 \cite{kay1993fundamentals}:
\begin{equation}\label{L1}
  \hat{\bs}_i=\underbrace{\bC_{\bs_i\by_i}\bC_{\by_i}^{-1}}_{\triangleq\bW_i} \by_i
\end{equation}
where  $\bC_{\bs_i\by_i}$ is the cross-covariance matrix between $\bs_i$ and $\by_i$ and $\bC_{\by_i}$ is the auto-covariance matrix of $\by_i$. Using \eqref{model} and noting that $\mathbb{E}\{\bs_i\bs_i^H\}=\bI_{d_i}$ and $\mathbb{E}\{\bs_i\bs_j^H\}=\textbf{0}, i\neq j$, we have:
\begin{eqnarray} \label{L2}
  \bC_{\bs_i\by_i}=\mathbb{E}\{\bs_i\by_i^H\}&=& \bV_i^H \bH_{ii}^H \\ \nonumber
  \bC_{\by_i} = \mathbb{E}\{\by_i\by_i^H\} &=& \sum_{j=1}^N \bH_{ji} \bV_j \bV_j^H \bH_{ji}^H + \bGamma_i
\end{eqnarray}
The expression of the LMMSE in \eqref{LMMSE} is obtained by substituting \eqref{L2} in \eqref{L1}.

 Next, we show that  substituting \eqref{LMMSE} in \eqref{Rate1} yields the expression for rate $R_i$ in \eqref{Rate}. To this end, we rewrite \eqref{LMMSE} by using the matrix inversion identity $(\bA+\bB\bC\bD)^{-1}\bB\bC=\bA^{-1}\bB(\bC^{-1}+\bD\bA^{-1}\bB)^{-1}$ as follows:
\begin{equation}\label{L3}
  \bW_i^{\textrm{LMMSE}}=\underbrace{\left(\bI_{d_i} + \bV_i^H \bH_{ii}^H  \bC_{\overline{i}}^{-1} \bH_{ii} \bV_i \right)^{-1}}_{\triangleq \bC_e} \bV_i^H \bH_{ii}^H \bC_{\overline{i}}^{-1}
\end{equation}
Let $\bOmega_i\triangleq \bW_i \bH_{ii} \bV_i \bV_i^H \bH_{ii}^H \bW_i^H \left(\bW_i \bC_{\overline{i}} \bW_i ^H\right)^{-1}$, then
\begin{align}\label{L4}
  \bOmega_i=&\bC_e \bV_i^H \bH_{ii}^H \bC_{\overline{i}}^{-1}  \bH_{ii} \bV_i \bC_e^{-1} \times \\ \nonumber
   &\bC_e \bV_i^H \bH_{ii}^H \bC_{\overline{i}}^{-1}\bH_{ii} \bV_i \bC_e  \left(\bC_e \bV_i^H \bH_{ii}^H \bC_{\overline{i}}^{-1}
  \bH_{ii} \bV_i \bC_e \right)^{-1} \\ \nonumber
  =& \bC_e \bV_i^H \bH_{ii}^H \bC_{\overline{i}}^{-1}  \bH_{ii} \bV_i \bC_e^{-1}
\end{align}
  Finally, it is readily verified that by substituting \eqref{L4} in \eqref{Rate1} and using  Sylvester determinant property, \eqref{Rate} is obtained.

  \section{Extension to MIMO-IBC }\label{Appendix B}
To model MIMO IBC, we consider $N$ cells that each consists of one transmitter (base station) and multiple receivers. More precisely, the transmitter of the $n$th cell ($n\in\{1,...,N\}$)  is equipped with $M_n$ antennas and serves $I_n$ receivers. Also, we denote the $i$th receiver in the $n$th cell by $i_n$ and its number of antennas by $L_{i_n}$.
The $n$th transmitter uses the linear precoder matrix $\bV_{i_n}\in \complexC^{M_n\times d_{i_n}}$ to send the information to its $i$th receiver. More precisely,  it uses $\bV_{i_n}$ to convert the symbol stream $\bs_i\in \complexC^{d_{i_n}\times 1}$  into the vector $\bd_{i_n}\in \complexC^{M_{i_n}\times 1}$, i.e.,
\begin{equation}\label{t_beamformer1}
  \bd_{i_n}=\bV_{i_n} \bs_{i_n}
\end{equation}
and sends  $\sum_{i=1}^{I} \bd_{i_n}$ over flat fading channels. The received signal at the $i_n$th receiver is given by:
\begin{equation}\label{model1}
  \by_{i_n}=\underbrace{\bH_{n{i_n}}\bd_{i_n}}_{\textrm{desired signal}}
  +\underbrace{\sum_{k\neq i}\bH_{n{i_n}}\bd_{k_n}}_{\textrm{intracell interference}}+
  \underbrace{\sum_{j\neq n}\sum_{k}\bH_{j{i_n}}\bd_{k_j}}_{\textrm{intercell interference}}
  +\bn_{i_n}
\end{equation}
where $\bH_{j{i_n}}\in \complexC^{L_{i_n}\times M_j}$ denotes the channel matrix between the $j$th transmitter and the ${i_n}$th receiver. Also, $\bn_{i_n}\in \complexC^{L_{i_n}\times 1}$ is the CSCG noise at the ${i_n}$th receiver with zero mean and covariance matrix $\bGamma_{i_n} \in \realS^{++}_{L_{i_n}}$.
The maxmin optimization problem can be cast as  follows:
\begin{align}\label{p11}
\!\max_{\{\bQ_{i_n}\}_{{i_n=1}}^{I_N}}\!\! \min_{{i_n}=1,2,\dots , I_N} \;\; & {R_{i_n}} \\ \nonumber
\text{s.t.} \quad& \sum_{i=1}^I \tr\{\bQ_{i_n}\} \leq  p_{{i_n}} \; & \forall {i_n}=1,2,\dots , I_N \\ \nonumber
& \bQ_{i_n} \succeq \textbf{0} \; &\forall {i_n}=1,2,\dots , I_N \nonumber
\end{align}
where,
 \begin{align}\label{Rate22}
  \!\!\!&R_{i_n}  = \\ \nonumber
\!\!\! & \log \det \!\left(\bI_{L_{i_n}}\!\!+\!\!\bH_{n{i_n}} \bQ_{i_n} \bH_{n{i_n}}^{H}\!\left[\bGamma_{i_n}+  \!\!\!\!\!\!\!\sum_{(j,k)\neq (n,i)}\!\!\!\!\!\!  \bH_{j{i_n}} \bQ_{k_j} \bH_{j{i_n}}^H\right]\!\!^{-1}\right)
\end{align}
 with $\bQ_{i_n}  \triangleq \bV_{i_n} \bV_{i_n} ^{H} \in \complexC^{M_{i_n} \times M_{i_n} }, n=1,\dots , N, i=1,\dots , I$, being the precoder covariance matrices. The design problem above is similar to that in (\ref{p}) and hence can be dealt with by the proposed MM method.

\section{Proof of proposition 1}\label{Appendix C}
First, note that $\bQ_i$  can be decomposed as $\bQ_i=\widetilde{{\bV}}_{i}\widetilde{{\bV}}_{i}^{H}$. Let $\bB_{i,11}$ denote the left upper block of $\bB_i^{-1}$.
 %\begin{equation}\label{B1}
%    \bB_i^{-1}=\left[
% \begin{array}{cc}
%   \overline{\bB}_{i,11} & \overline{\bB}_{i,12} \\
%   \overline{\bB}_{i,21} & \overline{\bB}_{i,22}
% \end{array}
% \right].
% \end{equation}
By using the blockwise matrix inversion lemma (see, e.g., \cite{stoica2005spectral}), we have that:
 \begin{equation}\label{Inv1}
   {\bB}_{i,11}\!\!=\!\! \left(\!\bI_{M_i}-\widetilde{{\bV}}_{i}^{H}\bH_{ii}^{H}\!\!
   \left[ \bGamma_i+\displaystyle\sum_{j=1}^{N}\bH_{ji}\widetilde{{\bV}}_{j}\widetilde{{\bV}}_{j}^H\bH_{ji}^{H}\right]^{-1}
   \!\!\!\!\!\!\bH_{ii}\widetilde{{\bV}}_{i}\!\right)^{-1}
 \end{equation}
 Then, by using Woodbury matrix identity, \eqref{Inv1} can be rewritten as:
  \begin{equation}\label{Inv2}
   {\bB}_{i,11}\!\!=\!\! \bI_{M_i}+\widetilde{{\bV}}_{i}^{H}\bH_{ii}^{H}\!\!\!
   \left[ \bGamma_i+\displaystyle\sum_{j\neq i}\bH_{ji}\widetilde{{\bV}}_{j}\widetilde{{\bV}}_{j}^H\bH_{ji}^{H}\right]^{-1}
   \!\!\!\!\!\!\bH_{ii}\widetilde{{\bV}}_{i}
 \end{equation}
 Finally, substituting \eqref{Inv2}  in \eqref{R1} and using Sylvester determinant property, \eqref{Rate2} is obtained.
 %\hfill $\blacksquare$

\section{Proof that $\bB_i\succ \bf{0}$}\label{Appendix D}
The matrix $\bB_i$ is defined in \eqref{bi}.
%\defff
%\normalfont Consider a matrix $\bX$ partitioned as$$\bX=\left(
%                                        \begin{array}{cc}
%                                          \bA & \bB \\
%                                          \bB^T & \bC \\
%                                        \end{array}
%                                      \right),$$
%                                       where $A\in \realS_K$. If $\det A\neq 0$, the matrix $\bS=\bC-\bB^T\bA^{-1}\bB $ is called the Schur complement of $\bA$ in $\bX$.
%\thm
%\normalfont Let $\bX=\left(
%                                        \begin{array}{cc}
%                                          \bA & \bB \\
%                                          \bB^T & \bC \\
%                                        \end{array}
%                                      \right)$ be a matrix and S be the Schur complement of $\bA$ in $\bX$. Then $\bX \succ \bf{0}$ if and only if $\bA\succ \bf{0}$ and $\bS\succ \bf{0}$ \cite{boyd2004convex}.
First it is obvious  that $\bI_{M_i}\succ\bf{0}$. Thus, it suffices to prove that the Schur complement of $\bI_{M_i}$ in $\bB_i$ is positive definite, i.e. \cite{boyd2004convex}\cite{stoica2005spectral}:
\begin{align}\label{111}
  \bS_i\triangleq &\bGamma_i+\displaystyle\sum_{j=1}^{N}\bH_{ji}\bQ_{j}\bH_{ji}^{H}-\bH_{ii}\bQ_i\bH_{ii}^{H} \\ \nonumber
  = &\bGamma_i+\displaystyle\sum_{\underset{j\neq i}{j=1}}^{N}\bH_{ji}\bQ_{j}\bH_{ji}^{H}\succ \bf{0} \nonumber
\end{align}
The matrices $\bH_{ji}\bQ_{j}\bH_{ji}^{H}, \forall i,j$ are obviously positive semidefinite. Therefore, $\bS_i$ is positive  definite because it is the sum of a positive definite matrix  ($\bGamma_i$) and a  number of positive semidefinite matrices, and as a result $\bB_i\succ{0}$.%\hfill $\Box$

\section{Proof of eq. \eqref{hyp}}\label{Appendix E}
%As stated in Lemma 1, $f(\bB_{i})\triangleq \log \det (\bU^{H}\bB_{i}^{-1}\bU)$ is convex with respect to $\bB_{i}$ and  can be underestimated by the  first order condition as \cite{boyd2004convex}:
%\begin{align} \label{hyp1}
% f(\bB_{i}) \,\geq\,& f(\overline{\bB}_{i})+ \tr\{\nabla^T_{\bB_{i}}f(\overline{\bB}_{i})(\bB_{i}-\overline{\bB}_{i})\},
%  \quad \forall \bB_{i},\overline{\bB}_{i}\succeq \textbf{0}
%\end{align}
%where $\nabla_{\bB_{i}}f(\bB_{i})$ is the gradient of $f(\bB_{i})$ with respect to the matrix $\bB_{i}$ and is given by \cite{petersen2008matrix}:
%\begin{align} \label{hyp2}
%\nabla_{\bB_{i}}f(\bB_{i})= -{\bB}_{i}^{-1}\bU(\bU^{H}{\bB}_{i}^{-1}\bU)^{-1}\bU^{H}({\bB}_{i}^{-1})^{H}
%\end{align}
%Finally, \eqref{hyp} is obtained by substituting \eqref{hyp2} in \eqref{hyp1}.  \hfill $\Box$
We begin the proof by presenting the following theorem from \cite{zalinescu2002convex}.
%We  prove \eqref{hyp} by means of the following theorem \cite{zalinescu2002convex}.
\thm{
\normalfont Let $\bX \in \realS^{+}_N$ and define
\begin{equation*}
  \nabla_{\bX}f(\bX)=\begin{bmatrix}
   \frac{\partial f(\bX)}{\partial \bX_{11}}       & \frac{\partial f(\bX)}{\partial \bX_{12}} & \frac{\partial f(\bX)}{\partial \bX_{13}} & \dots &\frac{\partial f(\bX)}{\partial \bX_{1N}} \\
    \frac{\partial f(\bX)}{\partial \bX_{21}}      & \frac{\partial f(\bX)}{\partial \bX_{22}} &\frac{\partial f(\bX)}{\partial \bX_{23}} & \dots &\frac{\partial f(\bX)}{\partial \bX_{2N}} \\
    \hdotsfor{5} \\
   \frac{\partial f(\bX)}{\partial \bX_{N1}}       & \frac{\partial f(\bX)}{\partial \bX_{N2}} &\frac{\partial f(\bX)}{\partial \bX_{N3}} & \dots & \frac{\partial f(\bX)}{\partial \bX_{NN}}
\end{bmatrix}
  \end{equation*}
  for a differentiable function $f(\bX): \realS^{+}_N\rightarrow\realR$.
 Then, the following inequality holds for  any convex (differentiable) function $f(\bX)$:
   %
%Let  $f(\bX): \realS^{++}_N\rightarrow\realR^{+}$ be a convex differentiable function. Then, the following statement holds:
\begin{align} \label{hyp10}
 \!\!\!\!\!\!\!\!\!f(\bY) \,\geq\,& f(\bX)+ \tr\{\left(\nabla_{\bX} f(\bX)\right)^H(\bY-\bX)\},
  \; \forall  \bX,\bY \succeq \textbf{0}
\end{align}
%Where $\nabla_{}f(\bX)=\lim_{\lambda\to0}\frac{f(X+\lambda H)-f(X)}{\lambda} \;\;    (\forall  H \succeq \textbf{0}) $.
}\hfill$\blacksquare$
\normalfont

Now, we use the following differentiation formulas for $g(\bX)\triangleq \det(\bA \bX^{-1} \bB)$ with $\bX \in \realS^{++}_N$ and $\bA$, $\bB$ of proper dimensions:
\begin{align} \label{1}
%\nabla_{\bX}(\bX^{-1})=-\bX^{-2}
&\nabla_{\bX}\left(g(\bX)\right)=\\ \nonumber &-
\det(\bA\bX^{-1}\bB)\bX^{-1}\bA^H(\bB^H\bX^{-1}\bA^H)^{-1}\bB^H\bX^{-1}
\end{align}
Putting $\bA=\bU^H$ and $\bB=\bU$ in \eqref{1} and applying the chain rule, we can write:

\begin{align}\label{A1}
 &\nabla_{\bX}( \log(g (\bX)))=\frac{1}{g(\bX)}\nabla_{\bX}(g(\bX)) \nonumber
 \\
&=-{\bX}^{-1}\bU(\bU^{H}{\bX}^{-1}\bU)^{-1}\bU^{H}{\bX}^{-1}.
\end{align}
The proof of \eqref{hyp} is completed  by using \eqref{A1} in  (\ref{hyp10}).
%\hfill $\Box$

\section{Proof of Theorem 2}\label{Appendix F}
Let $f(\bx)\triangleq \underset{i}{ \min}\;f_i(\bx)$ and $g(\bx,\bx_0)\triangleq \underset{i}{ \min}\;g_i(\bx,\bx_0)$. To prove Theorem 2, we will show that the conditions \textbf{A.1}-\textbf{A.3} stated in Theorem 1 are satisfied for the problem \eqref{p2}.

$f(\bx) $ is a continuous function, since it is the minimum of finite set of continuous functions. Therefore, the set $A=f^{-1}((-\infty,f(\bx_{0})])$ is closed and as a result $lev_{\leq f (\bx_0 )}f$, which is a closed subset of compact set $C$, is also compact \cite{rudin1976principles}.

Since $M$ is finite, there is an interval $\Lambda= [0,\lambda_{\max}]$ such that $\min_{i} \; f_{i}(\bx_{0}+\lambda \bd)=f_{\hat{i}}(\bx_{0}+\lambda \bd)$ and $\min_{i} g_{i}(\bx_{0}+\lambda \bd,\bx_{0})= g_{\hat{i}}(\bx_{0}+\lambda \bd,\bx_{0})$ $, \forall \lambda \in  \Lambda$.
From convexity of $f_{\hat{i}}(\bx)$, we have:
\begin{align} \label{12}
  f_{\hat{i}}(\bx_{0}+\lambda \bd) \geq f_{\hat{i}}(\bx_{0})+ \tr\left(\nabla_{\bX} f_{\hat{i}}(\bx_{0})\right)^T(\lambda \bd))
 \;,  \forall \lambda \in  \Lambda
\end{align}
On the other hand, from the definition of $g(\bx,\bx_0)$:
  \begin{align} \label{122}
&\limsup_{\lambda\to0}\frac{g(\bx_{0}+\lambda \bd,\bx_{0})-g(\bx_{0},\bx_{0})}{\lambda}= \\ \nonumber
&\tr\{\left(\nabla_{\bX} f_{\hat{i}}(\bx_{0})\right)^T( \bd)\} \nonumber
 \end{align}
Combining \eqref{12} and \eqref{122} leads to:
 \begin{align} \label{AA2}
&\limsup_{\lambda\to0}\frac{f(\bx_{0}+\lambda \bd)-f(\bx_{0})}{\lambda}\geq\\  \nonumber
&\limsup_{\lambda\to0}\frac{g(\bx_{0}+\lambda \bd,\bx_{0})-g(\bx_{0},\bx_{0})}{\lambda}
\end{align}
Also for every $\epsilon>0$  there is some $\alpha$ such that the following inequality holds for $ \lambda \leq \alpha,\lambda \in \Lambda$:
\begin{align}
f_{\hat{i}}(\bx_{0}+\lambda \bd) \leq f_{\hat{i}}(\bx_{0})+ \tr\{\left(\nabla_{\bX} f_{\hat{i}}(\bx_{0})\right)^T \lambda \bd\}+\epsilon\lambda
\end{align}
Therefore:
 \begin{align}
\frac{f(\bx_0+\lambda \bd)-f(\bx_0)}{\lambda}\leq\frac{g(\bx_0+\lambda \bd,\bx_0)-g(\bx_0,\bx_0)}{\lambda}+\epsilon
\end{align}
\begin{align} \label{AA1}
&\limsup_{\lambda\to0}\frac{f(\bx_0+\lambda \bd)-f(\bx_0)}{\lambda} \leq \\ \nonumber
&\limsup_{\lambda\to0}\frac{g(\bx_{0}+\lambda \bd,\bx_{0})-g(\bx_{0},\bx_{0})}{\lambda}
\end{align}
The proof is completed by combining \eqref{AA1} and \eqref{AA2}.

\section*{Acknowledgment}
\addcontentsline{toc}{section}{Acknowledgment}
The authors would like to thank Prof. Farid Bahrami, the Department of Mathematical Sciences, Isfahan University of Technology, for his helpful comments on the convergence analysis of the proposed method.

\bibliographystyle{IEEEtran}
\bibliography{IEEEabrv,mmnref}

\end{document}